\documentclass[sigconf]{acmart}

\usepackage{amsmath}
\usepackage{booktabs}
\usepackage{graphicx}
\graphicspath{{figures/}}
\usepackage[labelformat=simple]{subcaption}

\usepackage{makecell}
\usepackage{array}

\newcolumntype{L}[1]{>{\raggedright\let\newline\\\arraybackslash\hspace{0pt}}m{#1}}
\newcolumntype{C}[1]{>{\centering\let\newline\\\arraybackslash\hspace{0pt}}m{#1}}
\newcolumntype{R}[1]{>{\raggedleft\let\newline\\\arraybackslash\hspace{0pt}}m{#1}}

\usepackage{natbib}
\usepackage{rotating}
\usepackage{adjustbox}
\usepackage{float}
\usepackage{placeins}

\usepackage[absolute,showboxes]{textpos}

\usepackage{xurl}

\usepackage{svg}

\usepackage[]{todonotes} 
\usepackage{multirow}

\usepackage{csvsimple, listings}  

\usepackage[hyperfirst=false]{glossaries}
\glsdisablehyper
\makeglossaries
\newacronym[plural={VMs},shortplural={VMs}]{vm}{VM}{virtual machine}
\newacronym[plural={APIs},shortplural={APIs}]{api}{API}{application programming interface}
\newacronym[plural={SLAs},shortplural={SLAs}]{sla}{SLA}{service level agreement}
\newacronym[plural={QoSes},shortplural={QoSes}]{qos}{QoS}{quality of service}
\newacronym[plural={IaaSs},shortplural={IaaSs}]{iaas}{IaaS}{Infrastructure as a Service}
\newacronym[plural={building blocks},shortplural={BBs}]{bb}{BB}{building block}
\newacronym[plural={data centers},shortplural={DCs}]{dc}{DC}{data center}
\newacronym[plural={availability zones},shortplural={AZs}]{az}{AZ}{availability zone}
\newacronym[plural={TXs},shortplural={TXs}]{tx}{TX}{transmitted}
\newacronym[plural={RXs},shortplural={RXs}]{rx}{RX}{received}
\newacronym[plural={DRSs},shortplural={DRSs}]{drs}{DRS}{distributed resource scheduler}
\newacronym[plural={NUMAs},shortplural={NUMAs}]{numa}{NUMA}{non-uniform memory access}
\newacronym[plural={AIs},shortplural={AIs}]{ai}{AI}{artificial intelligence}
\newacronym[plural={MLs},shortplural={MLs}]{ml}{ML}{machine learning}
\newacronym[plural={ERPs},shortplural={ERPs}]{erp}{ERP}{enterprise resource planning}
\newacronym[plural={GPUs},shortplural={GPUs}]{gpu}{GPU}{graphics processing unit}
\newacronym[plural={CIs},shortplural={CIs}]{ci}{CI}{continuous integration}
\newacronym[plural={CDs},shortplural={CIs}]{cd}{CD}{continuous delivery}
\newacronym[plural={HA},shortplural={HA}]{ha}{HA}{high-availability}

\newcommand{\result}[1]{}

\definecolor{myred}{cmyk}{0, 0.7808, 0.4429, 0.1412}

\newcommand{\done}[1]{}

\usepackage{pifont}
\newcommand{\cmark}{\ding{51}}%
\newcommand{\xmark}{\ding{56}}%

\usepackage{xspace}
\newcommand{\etal}{\textit{et al.}~}
\newcommand{\eg}{\textit{e.g.,}\xspace}
\newcommand{\ie}{\textit{i.e.,}~}

\newcommand{\etc}{\textit{etc.}~}
\newcommand{\one}{({\em i})\xspace}
\newcommand{\two}{({\em ii})\xspace}
\newcommand{\three}{({\em iii})\xspace}

\makeatletter
\renewcommand{\paragraph}[1]{\vspace*{0.03in}\noindent{\bf #1.}\hspace{0.25ex \@plus1ex \@minus.2ex}}
\newcommand{\paragraphNoDot}[1]{\vspace*{0.03in}\noindent{\bf #1}\hspace{0.25ex \@plus1ex \@minus.2ex}}
\makeatother

\usepackage{arydshln}
\newcommand*\dhline{\specialrule{0pt}{1pt}{0pt}\hdashline[.4pt/3pt]\specialrule{0pt}{0pt}{2pt}}

\copyrightyear{2025}
\acmYear{2025}
\setcopyright{cc}
\setcctype{by}
\acmConference[IMC '25]{Proceedings of the 2025 ACM Internet Measurement Conference}{October 28--31, 2025}{Madison, WI, USA}
\acmBooktitle{Proceedings of the 2025 ACM Internet Measurement Conference (IMC '25), October 28--31, 2025, Madison, WI, USA}
\acmDOI{10.1145/3730567.3764480}
\acmISBN{979-8-4007-1860-1/2025/10}



\begin{document}

\title[The SAP Cloud Infrastructure Dataset: A Reality Check of Scheduling and Placement of VMs in Cloud Computing]{The SAP Cloud Infrastructure Dataset:\\A Reality Check of Scheduling and \\Placement of VMs in Cloud Computing}

\author{Arno Uhlig}
\orcid{0009-0006-6309-2696}
\affiliation{
  \institution{TU Dresden}
  \city{Dresden}
  \country{Germany}
}
\affiliation{
  \institution{SAP SE}
  \city{Dresden}
  \country{Germany}
}
\email{arno.uhlig@sap.com}

\author{Iris Braun}
\orcid{0009-0000-0900-2158}
\affiliation{
  \institution{TU Dresden}
  \city{Dresden}
  \country{Germany}
}
\email{iris.braun@tu-dresden.de}

\author{Matthias W\"ahlisch}
\orcid{0000-0002-3825-2807}
\affiliation{
  \institution{TU Dresden}
  \city{Dresden}
  \country{Germany}
}
\email{m.waehlisch@tu-dresden.de}

\begin{abstract}

Allocating resources in a distributed environment is a fundamental challenge.
In this paper, we analyze the scheduling and placement of virtual machines (VMs) in the cloud platform of SAP, the world's largest enterprise resource planning software vendor.
Based on data from roughly 1,800 hypervisors and 48,000 VMs within a 30-day observation period, we highlight potential improvements for workload management. 
The data was measured through observability tooling that tracks resource usage and performance metrics across the entire infrastructure. 
In contrast to existing datasets, ours uniquely offers fine-grained time-series telemetry data of fully virtualized enterprise-level workloads from both long-running and memory-intensive SAP S/4HANA and diverse, general-purpose applications.
Our key findings include several suboptimal scheduling situations, such as CPU resource contention exceeding 40\%, CPU ready times of up to 220 seconds, significantly imbalanced compute hosts with a maximum CPU~utilization on intra-building block hosts of up to 99\%, and overprovisioned CPU and memory resources resulting into over 80\% of VMs using less than 70\% of the provided resources.
Bolstered by these findings, we derive requirements for the design and implementation of novel placement and scheduling algorithms and provide guidance to optimize resource allocations.
We make the full dataset used in this study publicly available to enable data-driven evaluations of scheduling approaches for large-scale cloud infrastructures in future research.

\end{abstract}

\maketitle

 \definecolor{boxgray}{rgb}{0.93,0.93,0.93}
 \textblockcolor{boxgray}
 \setlength{\TPboxrulesize}{0.7pt}
 \setlength{\TPHorizModule}{\paperwidth}
 \setlength{\TPVertModule}{\paperheight}
 \TPMargin{5pt}
 \begin{textblock}{0.8}(0.1,0.04)
   \noindent
   \footnotesize
   If you refer to this paper, please cite the peer-reviewed publication: Arno Uhlig, Iris Braun, Matthias W\"ahlisch. 2025. The SAP Cloud Infrastructure Dataset: A Reality Check of Scheduling and Placement of VMs in Cloud Computing. 
   In \emph{Proceedings of the 2025 Internet Measurement Conference (IMC '25)}. ACM, New York, USA. https://doi.org/10.1145/3730567.3764480
\end{textblock}

\section{Introduction}
\label{sec:intro}

Efficient placement and scheduling of \glspl{vm} are critical to optimizing resource utilization and reducing energy consumption, ensuring a high-performance cloud computing system.
Cloud environments and their usage, however, grow, challenging the support of highly dynamic and potentially unpredictable workloads \cite{pr12030519, Belgacem2022}.
This intensifies the complexity when matching available resources with user demands.

Additionally, the diverse and heterogeneous pool of hardware resources, including specialized components such as high-performance RAM and GPUs, introduces significant complexity. 
Ensuring high availability through robust fault tolerance mechanisms and orchestrating efficient scheduling across geographically distributed environments further exacerbates these challenges.
Addressing these multifaceted issues requires a comprehensive approach that integrates workload modeling, resource characterization, and algorithmic efficiency.

The successful design of scheduling algorithms to place VMs depends heavily on sound models of realistic workloads.
Inefficient and ineffective placement and scheduling strategies can lead to resource contention situations which subsequently result in performance degradation of workloads \cite{MADNI2016173}.
Moreover, fragmentation of workloads on hypervisors can lead to increased operational costs and energy consumption.

In this paper, we provide a detailed understanding of common workloads of a large-scale enterprise cloud environment.
Our analysis is based on measuring 1,800 hypervisors and 48,000 VMs deployed at SAP, a multinational software company serving numerous customers enterprise resource planning services.
In contrast to common assumptions used in many studies, we show that workloads vary significantly in terms of resource usage and fragmentation, as well as resource contention over time. 

The primary objective of this study is to analyze scheduling in large-scale environments and outline limitations of current approaches, in particular within the widely-adopted OpenStack framework \cite{openstack_footprint_2023}.
Common schedulers, such as the OpenStack Nova scheduler, are designed to ensure load balancing, bin-packing, and resource availability, but may lack the flexibility and capabilities for diverse workloads and large-scale settings \cite{9740780}.
Our evaluation aims to determine the applicability of existing scheduling solutions in these settings and to identify gaps in meeting the requirements of the workload.
The results from this study will be used to explore and inform the development of novel schedulers that address the challenges imposed on large-scale cloud infrastructures.

Our key contributions are as follows.
\begin{enumerate}
  \item This study sheds light on cloud infrastructure challenges within a large-scale, real-world, production environment with diverse workloads ranging from lightweight to memory-intensive applications.

  \item We present a comprehensive dataset of up to 48,000 \glspl{vm} and 1,800 hypervisors within a single regional deployment and evaluate aspects such as infrastructure and workload sizing, resource allocations, and scheduling efficiency.

  \item We identify several inefficiencies and shortcomings in scheduling with vanilla OpenStack.

  \item The data and artifacts used within this study are made publicly available.
This is the first public dataset that includes workloads of VMs and has a high sampling resolution to reduce measurement errors.

\end{enumerate}

Prior work primarily focused on solutions using artificial workloads and simulations.
While useful for initial assessments, these datasets fail to accurately represent the complexity and variability of workloads in real-world environments.
Moreover, many approaches rely on synthetic data or datasets unavailable to the public.
This lack of transparency limits the reproducability and raises questions on the generalizability of the presented approaches.
Without realistic and real-world data, evaluating the effectiveness of scheduling solutions in practical applications becomes significantly challenging.
Addressing this research gap can help to advance research on scheduling strategies capable of meeting the requirements of large-scale cloud computing environments.

The remainder of this work is structured as follows.
In \autoref{sec:background}, we establish the background on scheduling in multi-datacenter cloud computing environments including the architecture of OpenStack and the Nova scheduler.
Next, we present an overview of a large-scale, real-world cloud environment that is being evaluated within this study and highlight its infrastructure, workload composition, and key metrics for subsequent scheduling efficiency analysis.
In \autoref{sec:methods}, we outline the tooling and measurement methods to systematically analyze the environment and establish the foundation for the results.
\autoref{sec:results} presents our measurement results of resource utilization, workload characteristics and scheduling efficiencies under real-world and productive conditions.
We compare with related in \autoref{sec:related-work}.
\autoref{sec:discussion} contextualizes our findings and discusses the generalizability of our results and outline opportunities for improvements.
Finally, we conclude and give an outlook in \autoref{sec:conclusions}.

We hope that the analysis provided in this study contribute valuable insight into and recommendations for scheduling strategies deployed in large-scale cloud computing environments.
We also hope that our public dataset can serve as a basis for further exploration, enabling research on workload characterization, optimization of scheduling services, improved resource utilization, and detection of resource contention situations.

\section{Background}
\label{sec:background}

In this section, we provide the necessary background about scheduling in cloud computing environments.
Moreover, we introduce the characteristics that relate to scheduling in OpenStack and, in particular, OpenStack Nova and its scheduler as used in the SAP cloud infrastructure.

\subsection{Cloud Scheduling}
\label{sec:cloud-scheduling}

The evolution of distributed systems has led to modern and large-scale cloud computing environments.
Cloud computing provides an on-demand self-service to a shared pool of resources and services accessible through network.
In particular, \gls{iaas} offers capabilities to provision virtualized resources, such as servers, storage, network, and more \cite{nist800_145}.
A server or \gls{vm} is a virtualized environment in which an operating system and applications can be run, isolated from the host~\cite{popek_goldberg_virtualized_architectures}.
In OpenStack-based cloud environments, a \textit{flavor} defines a predefined template of vCPUs, memory, and storage resources for a \gls{vm}.
\glspl{vm} are instantiated according to these flavors, which ensures standardized configurations across the infrastructure.
However, the requested resources are described to be higher than the actually consumed resource \cite{saxena_vm_prediction, malik_resource_utilization}. This is known as overcommitment or overprovisioning.
Hypervisors or compute nodes provide the environment to run \glspl{vm} and manage resource allocations and guarantee their execution \cite{popek_goldberg_virtualized_architectures}.
They can reside within a physical or virtualized unit.

\begin{figure}%
    \centering
    \includegraphics[width=0.5\textwidth]{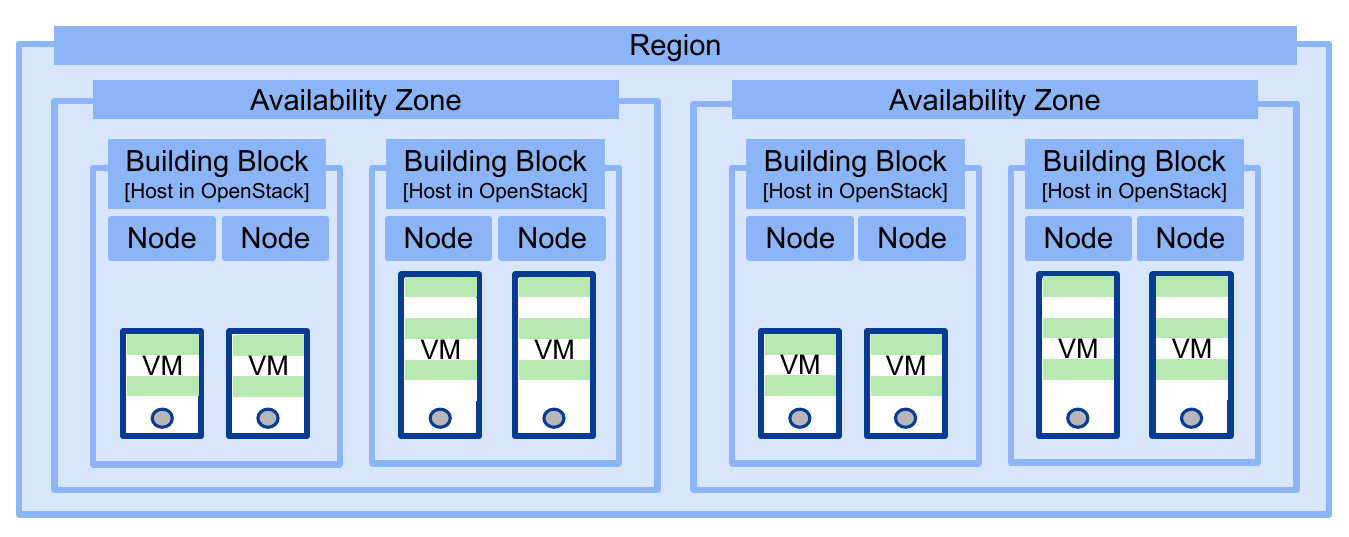}
    \caption{Hierarchical abstractions in cloud computing infrastructure}
    \label{fig:regional_architecture}
\end{figure}

The underlying infrastructure of cloud computing environments can be structured into several hierarchical components.
At the fundamental level are compute nodes and the physical hardware and infrastructure.
Multiple compute nodes can be grouped as a \gls{bb} for easier management and scalability effects.
A \gls{dc} provides the facilities to host multiple building blocks and provides support infrastructure such as power, cooling, and security.
A \gls{az} can be used to logically group a set of, potentially geographically co-located, but independent \glspl{dc} to ensure high-availability scenarios, increase resilience, and fault tolerance.
Regions are comprised of one or more \glspl{az} and represent the highest level in the presented hierarchy.
They are geographically distributed and enable customers to use resources and services that best fit their requirements.
\autoref{fig:regional_architecture} illustrates the different components.

Resource sharing needs to solve resource contention, which describes a situation in which multiple tasks or processes compete for the same shared resource. 
Scheduling allocates and manages a limited amount of system resources (\eg processing time, memory, and I/O devices) to processes to guarantee their execution.
This requires optimization of utilization, ensuring fair distribution while guaranteeing stability and performance demands.
Considering not only a single but a distributed and networked system, in which resources are spread across multiple, interconnected devices, the scheduler needs to cope with additional challenges.
Scheduling in a cloud infrastructure is complex as it needs to account for resource allocations, continuous load, and re-balancing within a dynamic, scalable, and heterogeneous environment while guaranteeing efficiency, minimal latency, and considering security.
Sub-optimal scheduling can cause over- or undercommitment of deployments, leading to too many or too little hardware resources.
This impairs the overall execution and might cause reduced performance.
Cloud providers still aim to efficiently manage (partly unpredictable) user demands and available resources.

\subsection{OpenStack and Nova Scheduler}
\label{sec:openstack}

OpenStack~\cite{openstack_what_is,openstack_nova_system_arch} is an open-source software framework to control the management of infrastructure resources such as compute, network, or storage throughout a data center.
It provides an alternative to hyperscalers such as Microsoft Azure, Google Cloud, and AWS, enabling private clouds.

\begin{figure}%
   \centering
   \includegraphics[width=0.5\textwidth]{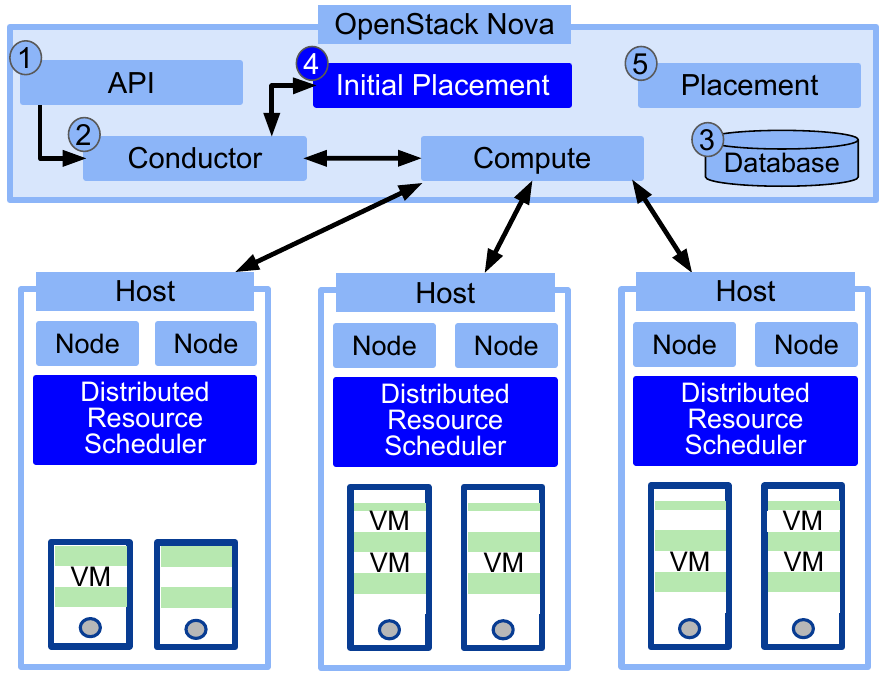}
   \caption{Simplified architecture of scheduling-relevant components in OpenStack Nova and VMware}
   \label{fig:nova_architecture}
\end{figure}

OpenStack was initially released in 2010 and the development was driven by Rackspace, NASA, and a majority of individual contributors.
The project received an increasing number of contributions and widespread adoption within the industry \cite{openstack_footprint_2023}. OpenStack can be considered the de facto standard within the context of private and hybrid cloud computing environments.
Several major organizations such as SAP drive and utilize OpenStack for large-scale cloud solutions.

OpenStack is modular following a service-oriented and distributed systems architecture.
A crucial component is Nova, the OpenStack compute engine, responsible for provisioning, managing, and scheduling of \glspl{vm}.
\autoref{fig:nova_architecture} illustrates the interplay of components relevant for scheduling using Nova. 
Existing dependencies to other OpenStack services such as Keystone (identity), Neutron (network), Glance (images), Cinder (block storage) are out of scope and not~shown.

In OpenStack, the term \textit{compute node} refers to a physical server running a hypervisor such as VMware ESXi.
A \textit{vSphere cluster} refers to an aggregation of compute nodes. 
This aggregation is represented as a \textit{compute host} in OpenStack.

The creation of a user-initiated \gls{vm} request within Nova involves multiple Nova sub-components.
A user specifies and requests a \gls{vm} via the \gls{api} as indicated by number (1) in the figure.
Subsequently, the request is handled by the Nova conductor~(2), which abstracts access to the Nova database~(3).
The scheduler~(4) is tasked with finding compute hosts that satisfy the resource requirements of the \gls{vm} while taking user and system constraints into account and ensuring optimal placement.
Despite being referred to as a \textit{scheduler}, the Nova scheduler does not perform continuous resource management.
Instead, it handles the placement of \glspl{vm} upon events triggered via the Nova API, such as \gls{vm} creation, resize, and migration.
Consequently, we refer to the scheduler as \textit{initial placement} in the illustration.
To make these decisions, it queries the placement API~(5), which maintains inventory and allocation records in its database.
This process is complemented by the Nova database, which persists relevant information.

\begin{figure}%
    \centering
    \includegraphics[width=0.35\textwidth]{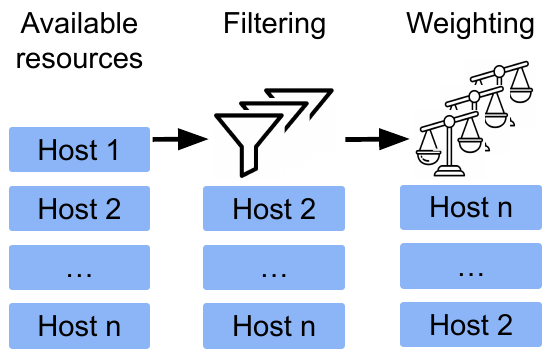}
    \caption{Scheduling of resources can be influenced by filtering and weighting, which increases complexity. Host numbering is reversed in the third process to illustrate how weighting can alter host priorities.}
    \label{fig:nova_scheduler}
\end{figure}

The Nova scheduler consists of a filter and weigher pipeline, see \autoref{fig:nova_scheduler}.
When receiving a compute request, the scheduler requests the list of all hypervisors.
Then, it applies a set of filters to eliminate hypervisors that do not meet the requirements of the requested \gls{vm}.
For example, the \textit{ComputeFilter} removes all hypervisors with insufficient compute resources (CPU, memory) for the \gls{vm}, the \textit{AvailabilityZoneFilter} ensures the \gls{vm} is assigned to the requested availability zone.
Subsequentially, weighers are used to generate a score and rank the remaining hypervisors in the list and provide the most suitable candidate.
Examples are the \textit{CPUWeigher} and \textit{RAMWeigher}, which prefer hypervisors with more resources.
Eventually, the \gls{vm} will be assigned to the highest ranked compute host.
The ranking process and the effect of weighting on priorities are illustrated in the third process by reversing the host numbering.
The set of pre-existing filters and weighers can be customized to influence scheduling decisions in order to meet user-specific requirements. 

The Nova scheduler does not support proactive resource reservation.
It favors dynamic use of maximal available resources rather than guaranteeing that a VM will always have access to sufficient resources when placed.
Instead, Nova implements a greedy approach with retries reapplying filters and weighers, which yields multiple suitable candidates.
The design of advanced scheduling behavior, VM~migration support \etc clearly benefit from detailed workload behavior.
Dynamic workload behavior is much more challenging to model, though.

\subsection{Value of our Experiences and Data}
\label{sec:literature-review}

Various studies (\eg \cite{6888900, 9339736}) review \gls{vm} scheduling strategies and present solutions to improve resource utilization in cloud computing environments.
They often rely on synthetic or simulated data.
Rana \etal~\cite{10.3389/fcomp.2024.1288552} provide a structured overview on current and emerging trends in the field of \gls{vm} scheduling, including the analysis of scheduling algorithms.
One major finding is that ``a major share of the literature studies is done on simulation-based tools using dummy datasets rather than real hypervisors'' \citep[p.~27]{10.3389/fcomp.2024.1288552}.
The authors emphasize that real-world data is important, as artificial data or simulations may fail to capture the dynamic, complex, and variable nature of real environments.
Moreover, synthetic data might fail to accurately capture anomalies, fluctuations in usage, patterns, correlations, unpredicted failures, and bottlenecks \citep{10.3389/fcomp.2024.1288552, Samsi_2021}.
We close this gap by providing data from a large-scale real-world deployment.

Gonzalez \etal\cite{Gonzalez2017} present a survey on resource management in cloud computing environments.
In particular, the lack of scalable resource management approaches capable of handling diverse workloads and potentially unpredictable demands is noted.
Moreover, the dynamic handling of variable compute and data requirements is identified as a challenge.
This clearly indicates the need for future work on resource management.
Our dataset, which we will make publicly available, will allow for realistic evaluation of such future work.

Sindhu \etal~\cite{Sindhu2022} highlight the importance of workload characterization to improve scheduling and capacity planning.
The authors propose a workload characterization model that is effective in dynamic environments.
While the authors claim the generated data could be representative of real-world workloads, its data model relies on the training set and workload characteristics.
Limitations of this approach are the workload-specific models, challenges with temporal changes and requirements for large, granular data.
While interesting, the work could benefit from large-scale and real-world datasets, particularly those with diverse and previously unknown characteristics, to validate and advance the approach.

\section{The SAP Cloud Infrastructure}
\label{sec:sap-cloud-overview}

In this section, we outline the SAP Cloud Infrastructure environment, including the deployment details.
The SAP Cloud Infrastructure is used by a wide range of internal business units and external partners within the regulated industry sector.
It encompasses more than 200,000 active \glspl{vm} and over 6,000 hypervisors distributed across 15 regions and 29 \glspl{dc}.
Our infrastructure includes more than 579,000 CPU cores and more than 16 petabyte of memory capacity.

In this paper, we focus on a regional deployment with 1,800 hypervisors and 48,000 VMs.  
This deployment was selected because it provides production-grade enterprise workloads suitable for analysis, while business confidentiality constraints required us to restrict the study to a single region.  
The dataset is representative of the broader SAP Cloud Infrastructure.

\subsection{Overview}
The SAP Cloud Infrastructure is the cloud computing offering of SAP.
It originates from the SAP Converged Cloud, a private managed cloud introduced in 2017.
Fundamentally, it is based on two building blocks:
Kubernetes, a widely adopted container orchestration platform, and OpenStack, an open-source cloud infrastructure framework.
It provides various services to manage an infrastructure, such as compute, network, object and block storage, network, and DNS.
Additional capabilities are provided by custom developed tools and services.

The SAP Cloud Infrastructure supports a variety of workloads, including enterprise and data analytic applications and \gls{erp} systems capable of real-time and large-scale data processing.
The platform is globally available with deployments in 29 \acrlongpl{dc} worldwide.

Our dataset captures real-world, production-grade enterprise workloads, including memory-intensive SAP S/4HANA systems.
These are representative of the landscape used for internal and customer-facing deployments.
Based on internal infrastructure data, 99.91\% of S/4HANA systems run in this environment, compared to 0.59\% on Microsoft Azure.
These enterprise workloads can be clearly distinguished from scientific or batch-oriented jobs in other environments due to their high memory demand, and stringent performance and availability requirements.

\paragraph{Data center architecture}
The SAP Cloud Infrastructure is based on a common hierarchy of regions, \acrlongpl{az}, \acrlongpl{dc}, \acrlongpl{bb}, and hosts (see \autoref{sec:background}).
To virtualize the data center infrastructure, the SAP Cloud Infrastructure uses VMware.
Each vSphere cluster represents a \acrlong{bb}, consisting of multiple VMware ESXi hypervisors (\ie hosts).
Multiple clusters may exist within an \acrlong{az}.
VMware vCenter serves as the management platform, which provides an interface for configuring and monitoring vSphere clusters and their hypervisors.

In Nova, each vSphere cluster is represented as a single compute host with respect to the placement \gls{api}.
Consequently, the Nova scheduler assigns a \gls{vm} to a specific cluster rather than to an individual hypervisor.
This abstraction can lead to fragmentation and imbalanced resource distribution situations within a vSphere cluster.
To mitigate this, the VMware \gls{drs} is used to dynamically balance \gls{vm}~workloads within a cluster.
The \gls{drs} is configured to monitor the load of the ESXi hosts and triggers automatic migrations of \glspl{vm} from over-utilized to less utilized hosts to ensure an optimal resource and load distribution.
Moreover, fragmentation and imbalances can also occur across \acrlongpl{bb}, requiring manual intervention or external rebalancers to resolve them.

\autoref{fig:regional_architecture} summarizes the data center deployment of two \acrlongpl{az}.
\autoref{fig:nova_architecture} highlights the components of the compute service involved in the creation, placement, and scheduling of \glspl{vm} within the OpenStack control plane.
Additional OpenStack components essential for \glspl{vm} such as Neutron (network), Glance (image), and Cinder (block storage) are not depicted.

\paragraph{Cross-datacenter migrations}
As mentioned, a regional deployment can consist of multiple, independent datacenters in geographical proximity considering infrastructure failure domains, such as the energy grid.
While this is required for \gls{ha} scenarios, migration of resource-intensive enterprise workloads beyond datacenter boundaries can involve additional technical and security-relevant considerations.
Within the context of this study, workload migrations across datacenter boundaries are out of scope and a single \gls{dc} is considered the placement and scheduling domain.

\paragraph{Support of high user demands}
To support user demands for high hardware resources, a subset of \acrlongpl{bb} is reserved allowing \gls{vm} flavors with special requirements such as \gls{gpu} workload and more than 3TB of memory.
These special purpose \acrlongpl{bb} do not accommodate other \glspl{vm} because special scheduling objectives may apply.
For flavors with at least 3TB of memory, the number of placeable \glspl{vm} is maximized to improve the utilization of the underlying compute hosts.

\paragraph{Geographical deployment}
Within the context of this work, the deployment of hypervisors and \glspl{vm} per data center represents the problem domain and well-defined boundaries for scheduling.
Cross-\acrlong{dc} scheduling and migrations are not considered.
This would require taking additional factors into account such as high network-related costs, latency, consistency, and fault-tolerance \cite{hogade2021energynetworkawareworkload}.

\begin{figure}%
    \centering
    \includegraphics[width=0.45\textwidth]{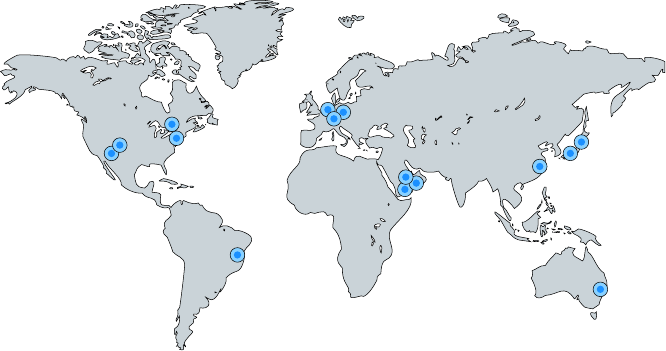}
    \caption{Regional deployments of the SAP Cloud Infrastructure across SAP owned and shared datacenters.}
    \label{fig:dc_worldmap}
\end{figure}

\autoref{fig:dc_worldmap} illustrates the various regional deployments of the SAP Cloud Infrastructure.
Each marker represents a \acrlong{dc} in Europe, North and South America, Asia, or Australia.
Customers can choose a region that best meets their requirements, reducing latency and maximizing performance.
All \acrlongpl{dc} are interconnected.

Each region consists of up to two data centers.
Depending on the region, a data center hosts 22 to 1072~hypervisors, providing capacity of up to 34,392~\glspl{vm}.
We present a detailed overview of the number of hypervisors and virtual machines per data center in \autoref{sec:dcs-details}.

Building block sizes range from 2 to 128 active compute nodes, depending on the region and observation period.
Smaller \glspl{bb} can lead to fragmentation, limit opportunities for improving resource utilization, and introduce additional challenges for scheduling.
Increasing \glspl{bb} sizes would be feasible and may reduce resource fragmentation. 
The rationale behind the current building block sizes is outside the scope of this work.

\subsection{Optimization Criteria and Constraints}
\label{sec:optimization-criteria}

Several optimization criteria and constraints are relevant for scheduling.
Our main objectives are to maximize the number of placeable \glspl{vm} per flavor, minimize fragmentation, optimize the overall resource utilization, and avoid resource contention situations.
Scheduling criteria that are highly specific to environment and workload are not discussed within this work.
Achieving the optimization criteria remains a challenge, though.
We now briefly discuss those challenges to relate later our empirical data to those challenges.

\paragraph{Assigning \glspl{vm} to host}
Bin packing can be used to optimize resource distribution and maximize the number of placeable \glspl{vm}, \ie fully utilize eligible hosts up to their configured limits before activating additional ones.
Bin packing is a classic NP-hard optimization problem that describes the process of fitting items into the minimal number of finite bins.
Well-known strategies with low computational effort include First-Fit, Best-Fit, and Worst-Fit.
These techniques have been extensively researched and various approximations can be leveraged to achieve our scheduling objectives~\cite{coffman_binpacking_1996, electronics7120389}.
However, challenges arise due to the dynamics of \gls{vm} sizes and the variability in host capacities. 
While hosts exhibit homogeneous hardware capabilities within a given \acrlong{bb}, they can differ between the \acrlongpl{bb} within an \acrlong{az}.
This introduces further complexities in optimal placement as the \acrlong{bb} specifics need to be considered.
Additionally, on-demand fluctuations can lead to frequent \gls{vm} migrations, which may degrade performance and increase operational costs.

During initial placement, the OpenStack Nova scheduler considers available CPU and memory resources, hardware compatibility, tenant isolation, host status, as well as VM and flavor specifications. 
The default strategy aims to load-balance general-purpose workloads, whereas SAP S/4HANA workloads are explicitly bin-packed to maximize memory utilization. 
While additional key performance indicators and metrics have since been integrated, they were not available at the outset. 
Changes to production systems require careful coordination and are subject to operational constraints.

\begin{figure*}[h!]%
    \centering
    \includegraphics[width=1.15\textwidth]{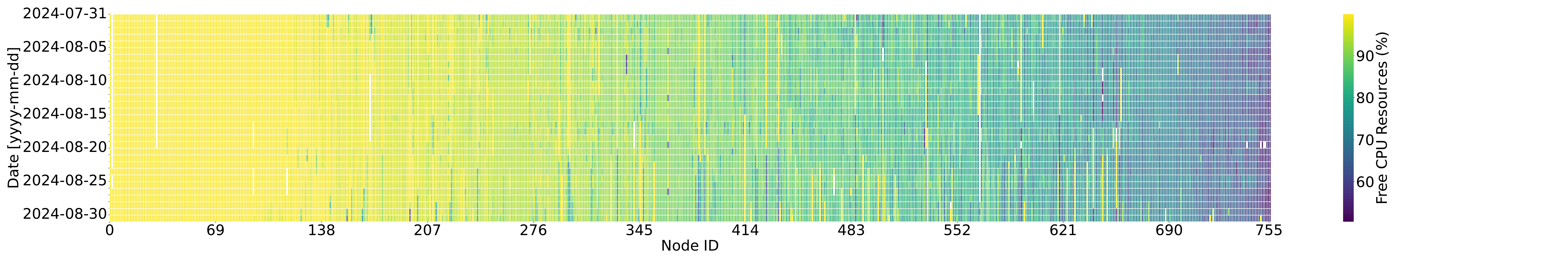}
    \caption{Daily average percentage of free CPU resources per host within a single data center}
    \label{fig:compute_host_cpu_usage_capacity}
\end{figure*}

\paragraph{Avoiding migration of heavy VMs}
Migrating \glspl{vm} that exhibit high CPU or memory operations should be avoided as migration results in overhead and performance degradation when large datasets are moved.
For example, a \gls{vm} maintains frequently changing in-memory data.
When implementing a seamless migration, either the (updated) memory~pages or deltas need to be copied from the original source to the new destination, or, alternatively, the migrated \gls{vm} reads the memory pages from the prior location.
Different solutions exist for both options but introduce performance penalties.
Therefore, it is preferred not to migrate but provide enough resources in advance.

\paragraph{Noisy neighbors}
The ``noisy neighbor'' effect describes the scenario when \glspl{vm} on the same host negatively impair each others' performance.
While bin packing can be an effective strategy to reduce the number of active hosts, distribution of resource-intensive workloads or workloads competing for similar resources on the same host remains an open problem.
In the SAP Cloud Infrastructure deployment, we try to avoid performance bottlenecks and local contention for shared resources, such as CPU, memory, and network bandwidth. 

\section{Data Collection and Processing}
\label{sec:methods}

In this section, we present the data collection process that creates the input for our analysis of the SAP Cloud Infrastructure usage.
We will make this dataset publicly~available.

Observability is crucial when operating a large-scale environment.
We regularly measure statistics that quantify the resource allocations of the different cloud components.
The monitoring backend for metrics at the SAP Cloud Infrastructure is based on Prometheus~\cite{prometheus}, an open-source monitoring system and time-series database.
To enable long-term, consolidated storage capabilities, Thanos~\cite{thanos} extends the monitoring system on top of Prometheus.
Both software platforms are industry standards with widespread adoption.
To query relevant statistics from the monitoring system and expose them towards Prometheus in a standardized format, common exporters are used~\cite{prometheus_instrumenting_and_exporters}.
The vROps exporter~\cite{sapcc_vrops_exporter} collects a comprehensive set of data from the VMware vRealize operations manager and provides a pre-defined set of metrics.
Another set of  metrics is directly extracted from the Nova database via the MySQL server exporter~\cite{prometheus_mysql_exporter}.
To distinguish both vROps and Nova data in a consolidated set, we use the specific name prefixes \textit{vrops} and \textit{openstack\_compute}.

Our dataset provides infrastructure resource utilization data collected at granularities ranging from 30 to 300 seconds.
Sampling intervals are defined by the respective component and represent a trade-off between measurement accuracy and observability overhead in our large-scale environment.
The telemetry data includes CPU and memory usage for each \gls{vm}, as well as scheduling-relevant events (if occurring within the observation period), such as creation, migration, resize, and deletion and enables a detailed analysis of scheduling decisions and resource contention.

All measurements and observations presented in this paper are based on the SAP Cloud Infrastructure monitoring system, which is used in operational deployment.
All observations were collected on July 31, 2024 00:00:00~(UTC+0) over the course of 30~days.
\autoref{sec:metrics} provides an overview of relevant metrics used in this study.

\paragraph{Dataset}
All artifacts, including software and raw data, are publicly available on Zenodo at \href{https://doi.org/10.5281/zenodo.17141306}{https://doi.org/10.5281/zenodo.17141306}~\cite{zenodo17141306}.

\begin{figure}%
    \centering
    \includegraphics[width=0.5\textwidth]{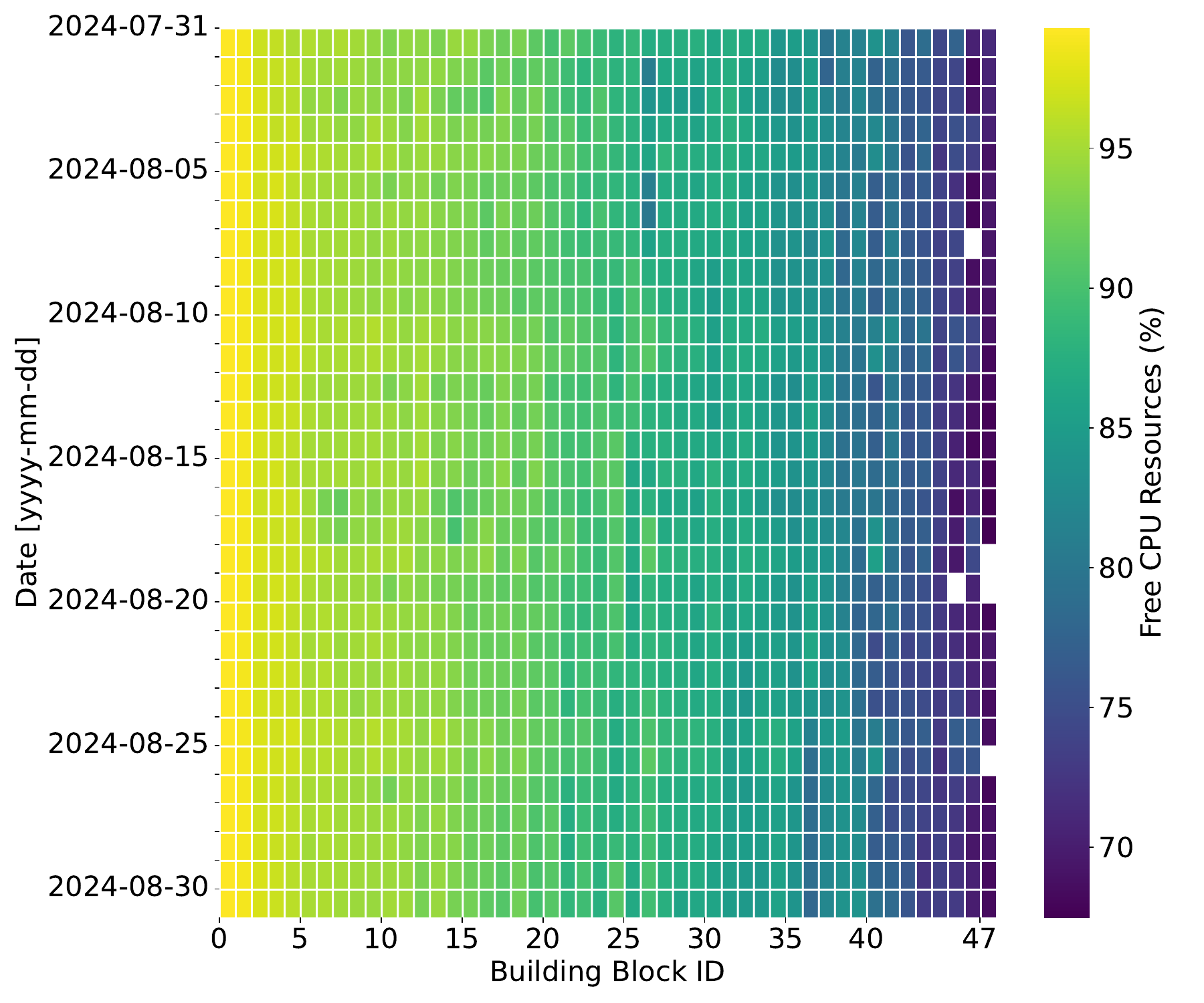}
    \caption{Daily average percentage of free CPU resources per building block in a single data center}
    \label{fig:building_block_cpu_usage_capacity}
\end{figure}

\section{Results}
\label{sec:results}

This section presents an analysis of the data collected within the SAP Cloud infrastructure.
We present the results along the different resources, CPU, memory, network, and storage, and conclude with an analysis of the workload composition.

\paragraph{Terminology}
Throughout this section, the term \textit{compute node} or \textit{node} refers to individual physical machines running hypervisors such as VMware ESXi.

A \textit{compute host} or \textit{host} refers to an aggregation of uniform compute nodes and is synonymous with a \textit{Building Block (BB)} and \textit{vSphere cluster}.

\paragraph{Understanding the Heatmaps}
The heatmaps in this section illustrate the average resource utilization within the observation period.
Each row shows a day within the considered period and a column corresponds to a compute host or \gls{bb} as per label and accompanying description.
Lighter colors indicate more free resource and darker colors a higher utilization and less free resources.
Moreover, compute hosts are sorted left to right from most to least free CPU resources.
White cells indicate missing data as compute hosts might have been added or removed during the observation period or experienced operational changes \eg planned maintenance.

\subsection{CPU}

\autoref{fig:compute_host_cpu_usage_capacity} presents the percentage of free CPU resources per compute node within a data center over 30 days.
Each row shows a day within the considered period and a column corresponds to a compute node.

The color gradients vary significantly between some compute node which highlights varying degrees of CPU utilization.
While some nodes are considerably utilized with less than 20\% free resources, other nodes show quite the contrary with 90\% or more free resources at the same day.
This observation is valid for the entire period.
A subset of nodes show consistent high CPU availability, suggesting underutilization, while others remain at low levels with potential resource contention situations.
Furthermore, some nodes show notable fluctuations over the observed period while others maintain their utilization levels.
This imbalanced distribution of workload within the data center may be intentional depending on the chosen scheduling algorithm.

Over the course of the 30-day period, some nodes show a consistent increase in CPU demand and less free resources.
This could be caused by an increasing demand of the existing workload or new \glspl{vm} being scheduled to the respective compute nodes.

The widespread underutilization of compute nodes suggests an overprovisioned data center.
This conclusion, however, might be misleading for several reasons.
\one In our environment, we aim to maximize the number of placeable \glspl{vm} of specific flavors on nodes in designated \glspl{bb} through bin-packing techniques.
This leads to strong separation of heavily utilized and less utilized compute hosts.
\two Capacities are intentionally reserved in case of emergency failover, redundancy, and scalability demands. 
\three The current situation allows considering additional constraints (see \autoref{sec:optimization-criteria}).
To better understand potential inefficiencies in workload and thus resource distribution, additional monitoring and rebalancing capabilities are needed.

\begin{figure}%
    \centering
    \includegraphics[width=0.5\textwidth]{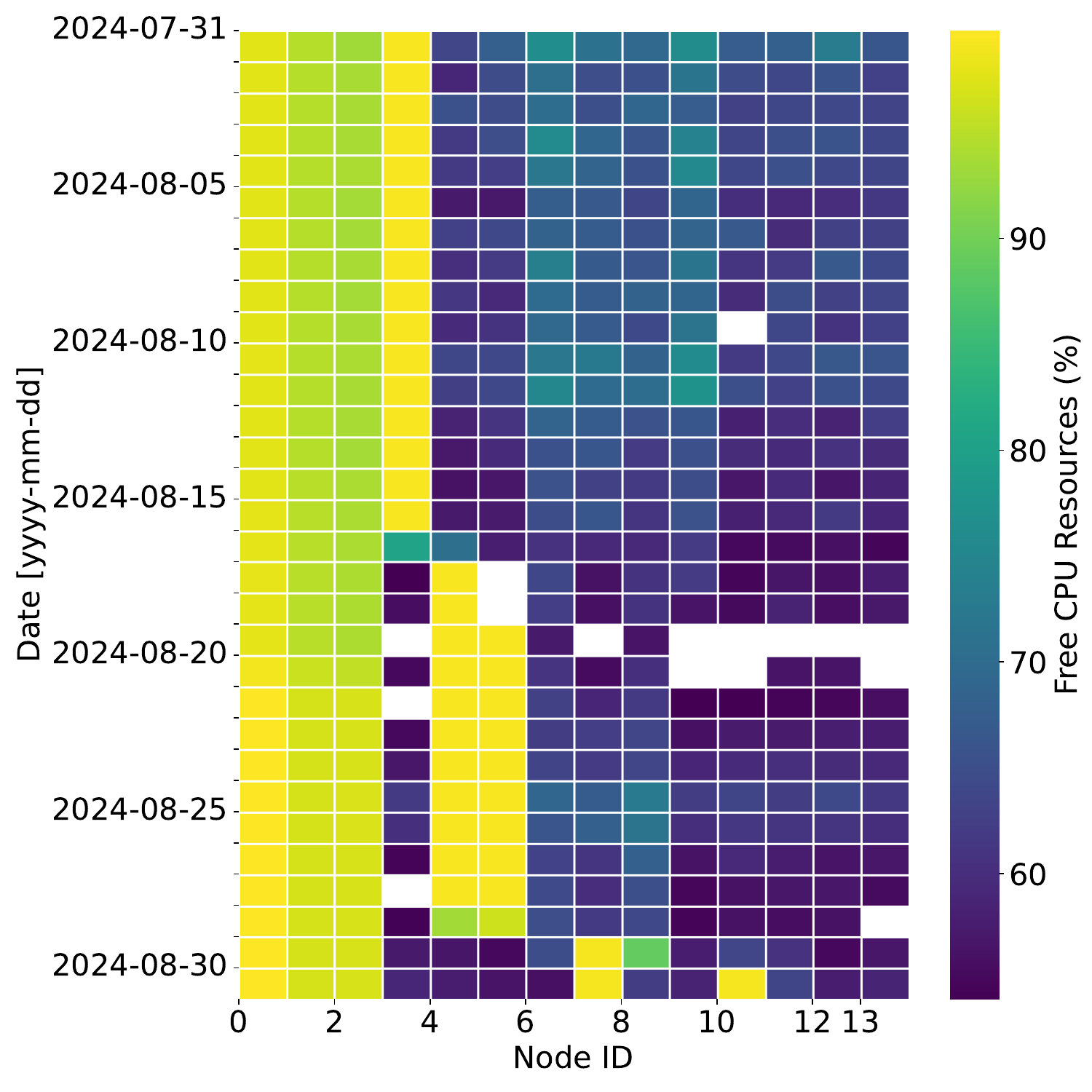}
    \caption{Daily average percentage of free CPU resources per node within a building block in a single data center}
    \label{fig:building_block_bb184_cpu_usage_capacity}
\end{figure}

\autoref{fig:compute_host_cpu_usage_capacity} provides a complete view on the CPU~utilization of our infrastructure.
We now zoom into different data center layers involved in scheduling.
\autoref{fig:building_block_cpu_usage_capacity} shows the results for all \acrlongpl{bb} within an \acrlong{az}, and \autoref{fig:building_block_bb184_cpu_usage_capacity} shows the utilization of CPU resources per compute node of a selected \acrlong{bb}.
Within a \acrlong{bb}, we can observe different levels of utilization. 
Some nodes are heavily utilized, while others show a significant amount of free resources.
Furthermore, several significant resource contention situations are visible.

In this paper, \textit{CPU contention} refers to time a virtual CPU (vCPU) is ready to execute instructions but cannot be scheduled on a physical CPU (pCPU).
This corresponds to the VMware metrics described in \autoref{sec:metrics}.
For example, a value of 40\% indicates that a vCPU waited 40\% of the observed time and could not execute instructions immediately.  

\begin{figure}%
    \centering
    \includegraphics[width=0.5\textwidth]{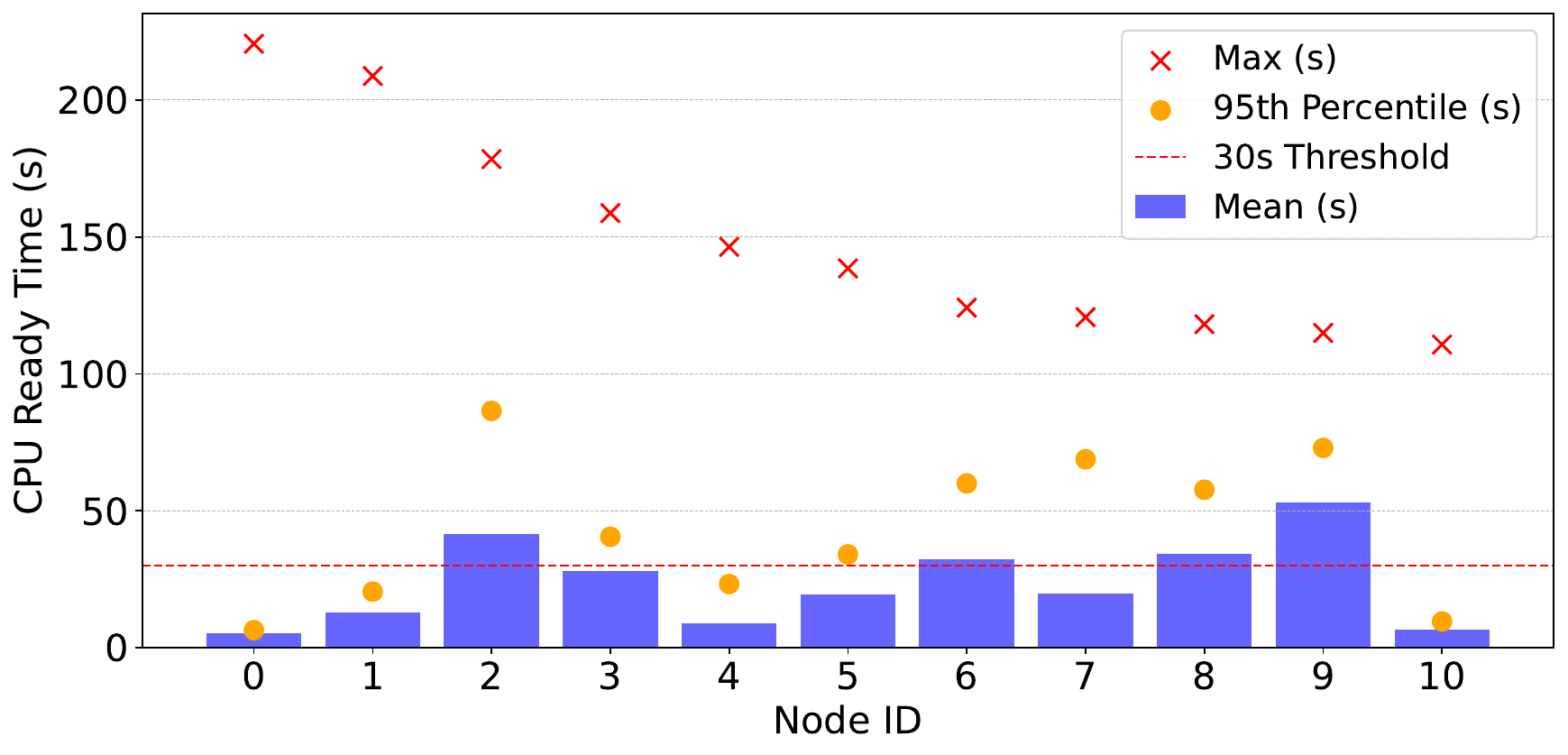}
    \caption{Aggregated CPU ready time of the 10 nodes with the highest CPU ready time across within the region.}
    \label{fig:cpu_ready_seconds}
\end{figure}

To understand contention situations in more detail, we measure the CPU ready time (\ie duration in which a \gls{vm} is ready to execute instructions but has to wait to get physical CPU resources) and the relative amount of contention per compute node.
\autoref{fig:cpu_ready_seconds} illustrates the 10 nodes with the highest CPU ready time across all datacenters of the region during our measurement period.
Multiple spikes can be observed throughout the entire month.
Outliers of around 30~minutes in the beginning of August might be indicative of exceptional situations.
However, the available data does not allow for a more detailed explanation.
Some temporal effects can be observed and indicate less workload and thus less contention on weekends and more during the working days.
The overall resource contention, however, remains noticeable.
Various hypervisors exceed the 30~second baseline several times throughout the month.

\begin{figure}%
    \centering
    \includegraphics[width=0.5\textwidth]{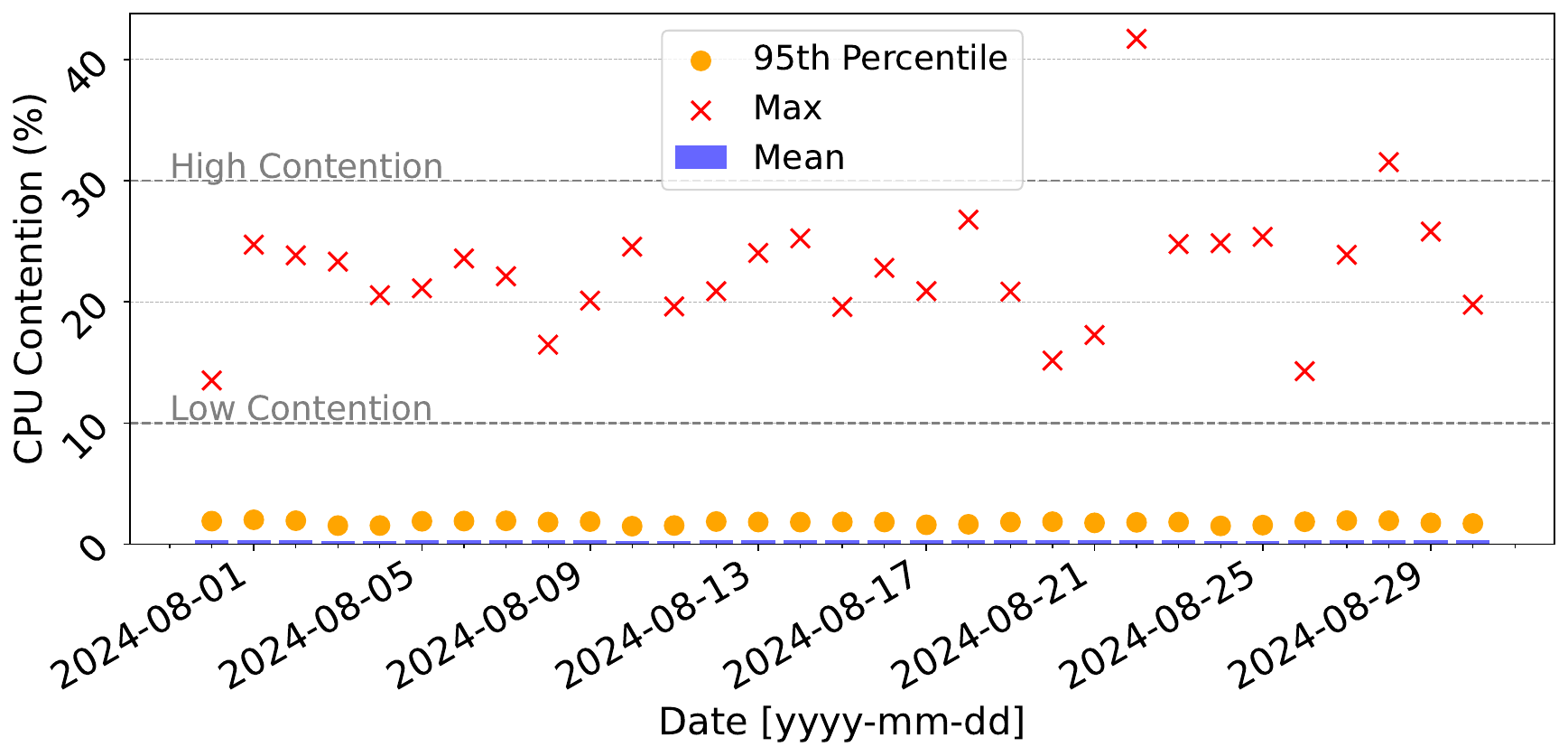}
    \caption{Aggregated CPU contention over all nodes within the region}
    \label{fig:cpu_contention_percentage_all}
\end{figure}

The CPU contention metric per compute node, as illustrated in \autoref{fig:cpu_contention_percentage_all}, shows increased levels of overutilization throughout the observed period.
While the daily mean and 95 percentile remain below the 5\% mark, the maximum contention of various nodes ranges between 10\% and 30\%.
The values exceed the strict threshold of 10\%, which is typically applied to critical workloads, but remain below the moderate threshold of 30\%, considered acceptable for time-sensitive systems.
However, several nodes exceed the 40\% level, indicating significant CPU contention situations.
We observe outliers of up to 40\% around mid-month, significantly saturating the CPUs.
The data is consistent across the observed period and does not show temporal effects, implying a persistent problem.

\subsection{Memory}

\begin{figure*}%
    \centering
    \includegraphics[width=1.15\textwidth]{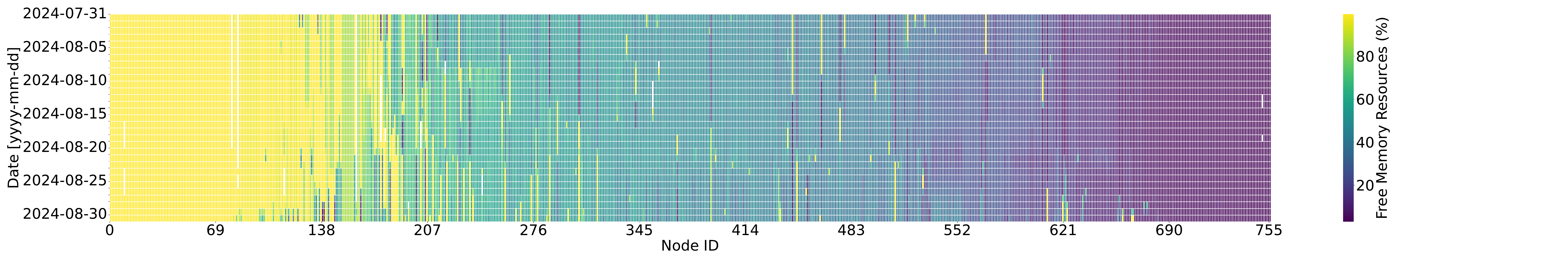}
    \caption{Daily average percentage of free memory resources per node within a single data center}
    \label{fig:compute_host_memory_usage_capacity}
\end{figure*}

Memory is a critical constraint when placing \glspl{vm}. %
This is especially relevant when extensive and low-latency memory access is required.
\autoref{fig:compute_host_memory_usage_capacity} illustrates the percentage of free memory.
The color gradient reflects more free resources (yellow) or less free resources (purple).
Compute nodes are sorted left to right from most to least free resources.

On the left side of the figure, nodes are visible that do not require much memory as indicated in yellow and green.
Roughly the same amount of compute nodes show less than 20\% of free memory and can be considered almost fully utilized.
We discuss more details in \autoref{sec:workload-composition}.

Some compute nodes show a slow increase in memory utilization and less free resources while other show consistent behavior per day.
Furthermore, significant and abrupt shifts from high to low memory utilization, as indicated by the color change from purple to yellow, can be observed.
These are caused by \gls{vm} migrations, shutdowns, or terminations.

The visual patterns indicate that some hosts become constrained in memory if the observed memory utilization patterns continue.

\subsection{Network}

\begin{figure*}%
    \centering
    \includegraphics[width=1.15\textwidth]{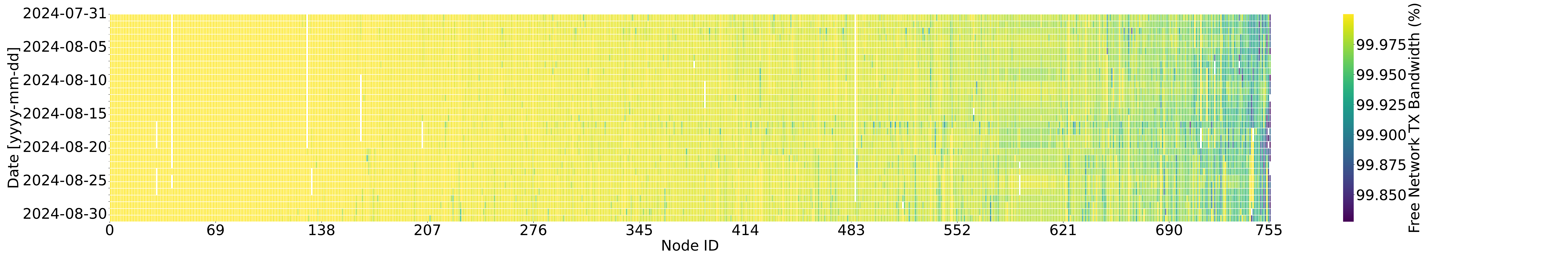}
    \caption{Daily average percentage of free network TX bandwidth per node within a single data center}
    \label{fig:compute_host_network_tx_usage_capacity}
\end{figure*}

\begin{figure*}%
    \centering
    \includegraphics[width=1.15\textwidth]{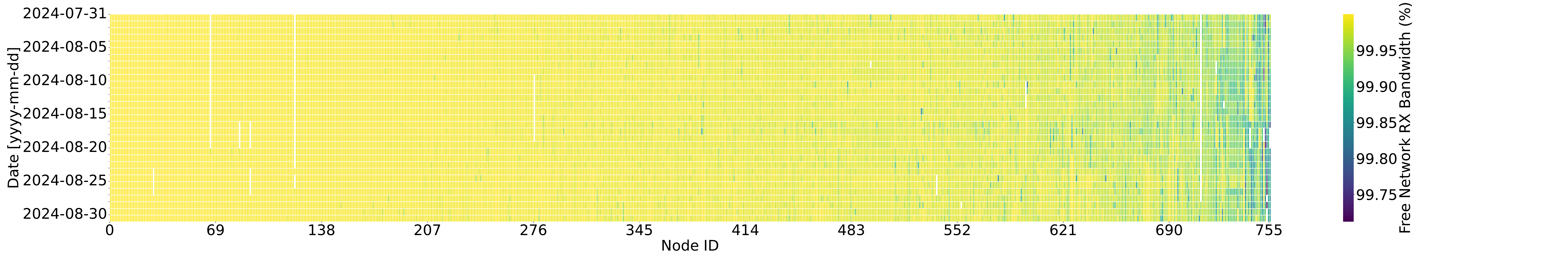}
    \caption{Daily average percentage of free network RX bandwidth per node within a single data center}
    \label{fig:compute_host_network_rx_usage_capacity}
\end{figure*}

\begin{figure*}%
    \centering
    \includegraphics[width=1.15\textwidth]{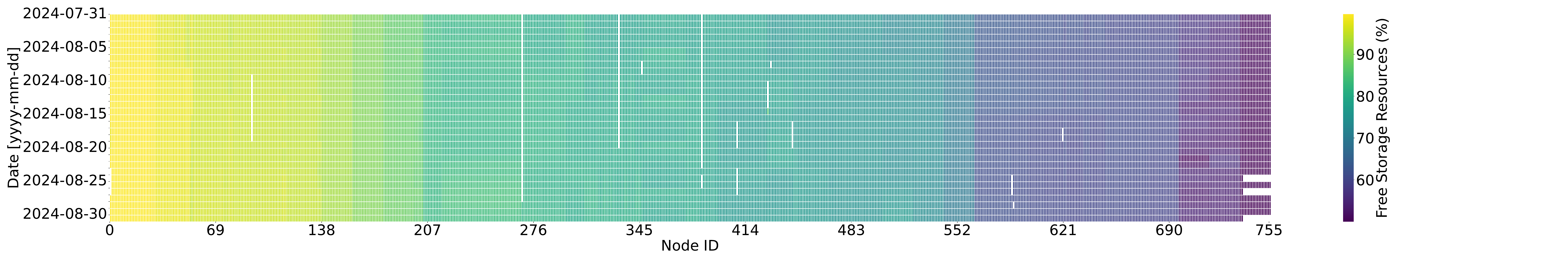}
    \caption{Daily average percentage of free storage resources per node within a single data center}
    \label{fig:compute_host_storage_usage_capacity}
\end{figure*}

Heavily utilized network interfaces of compute nodes could indicate network congestion and should be considered in scheduling decisions.
\autoref{fig:compute_host_network_tx_usage_capacity} and \autoref{fig:compute_host_network_rx_usage_capacity} show the percentages of \gls{tx} and \gls{rx} data via the network interfaces of compute nodes.
Each column represents a node and each row a day within the observed period.
The color gradient reflects the utilization of the network interface of each node.
Each compute node in the data center supports a maximum bandwidth of 200~Gbps.
It is clearly visible that the network load is notably below the 200~Gbps.
In the current deployment, considering network resources seems less relevant for scheduling decisions.

\subsection{Storage}

\begin{figure*}[htb]
    \centering
    \begin{subfigure}[b]{0.48\textwidth}
        \centering
	\includegraphics[width=\textwidth]{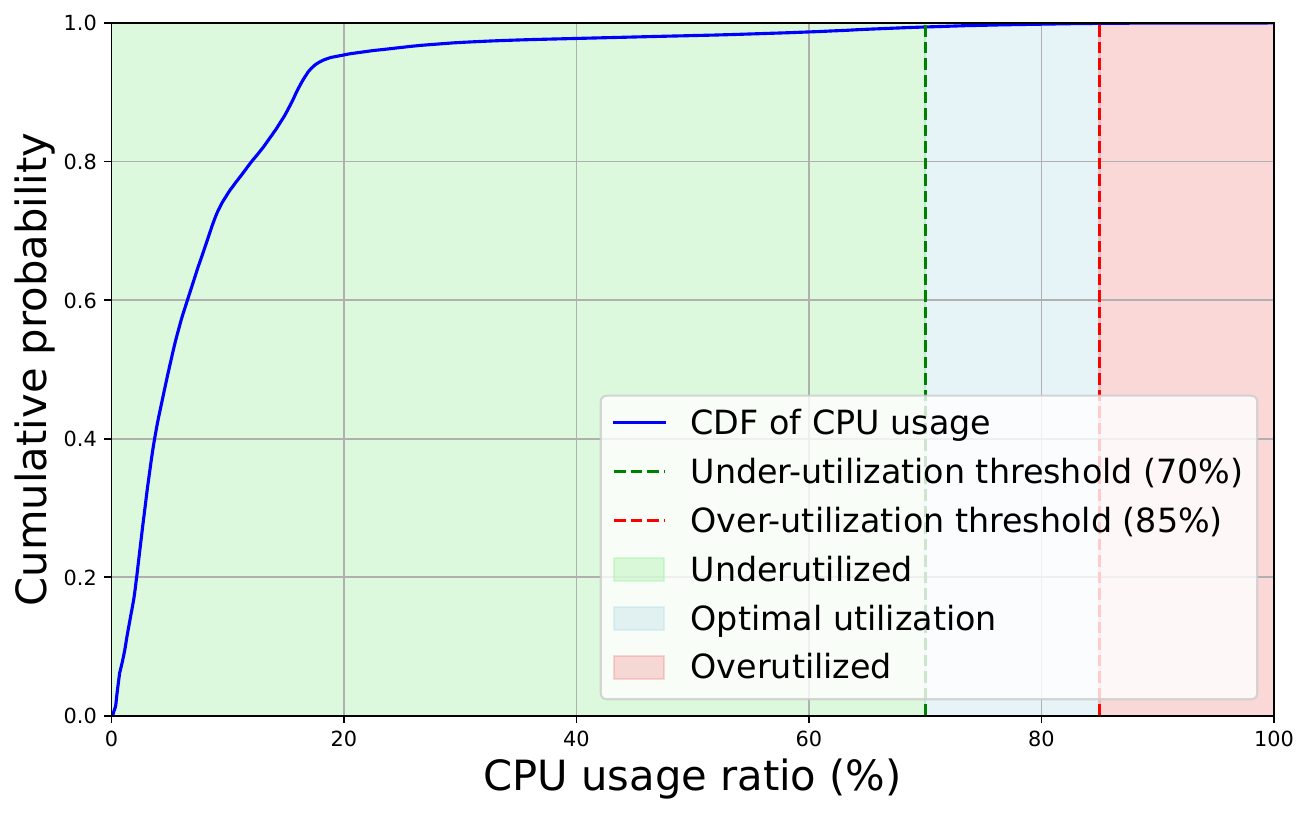}
        \caption{CPU resources: Virtual machines are predominantly overprovisioned in terms of CPU resources}
        \label{fig:vrops_virtualmachine_cpu_usage_ratio_all}
    \end{subfigure}
    \hspace{0.02\textwidth}
    \begin{subfigure}[b]{0.48\textwidth}
        \centering
	\includegraphics[width=\textwidth]{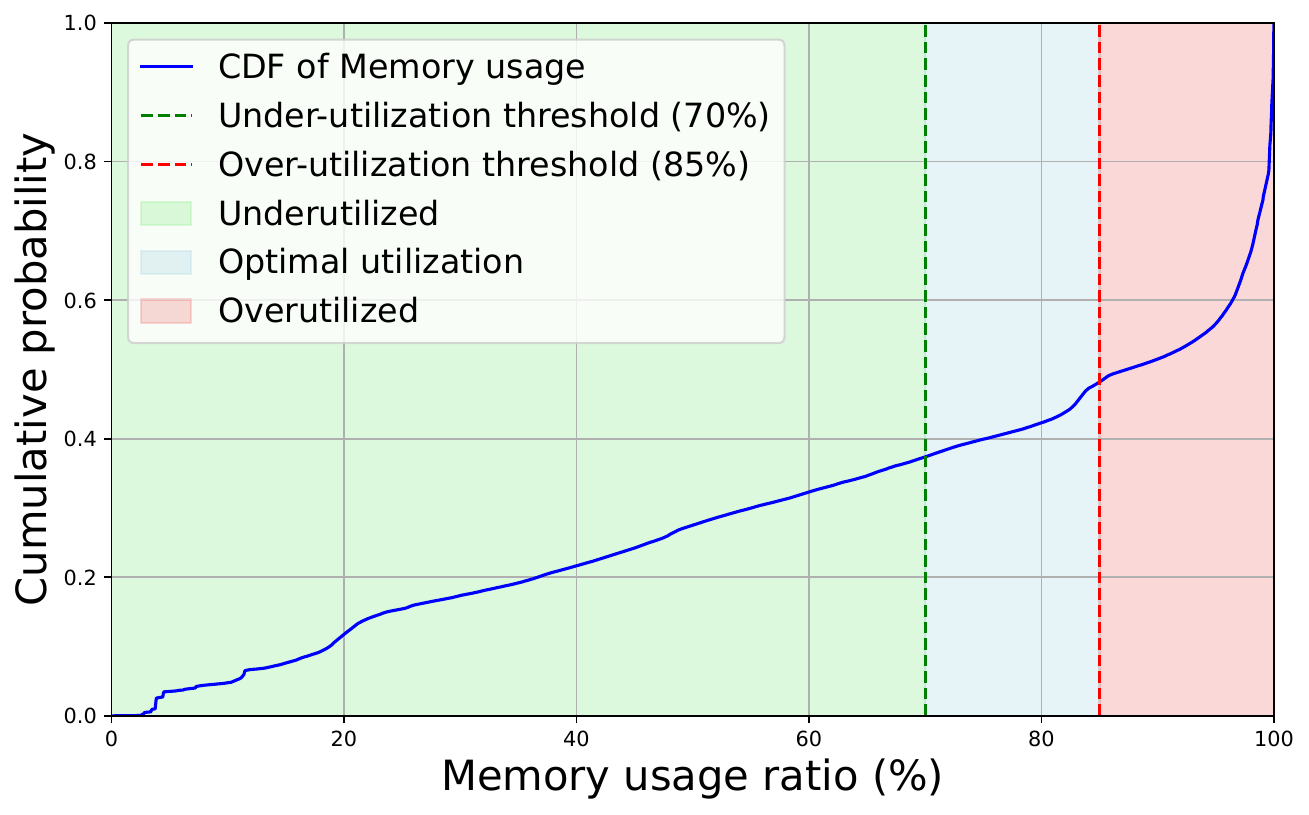}
        \caption{Memory resources: Virtual machines are often underprovisioned in terms of memory resources}
	\label{fig:vrops_virtualmachine_memory_usage_ratio_all}
    \end{subfigure}
    \caption{Cumulative distribution of average VM utilization ratio per resource}
    \label{fig:vm_resource_utilization_ratio}
\end{figure*}

In the SAP Cloud, the utilization of the local storage provided by each compute host is not considered with regard to scheduling.
This may provide opportunities to improve performance, fault tolerance and recovery, or cost efficiency.

\autoref{fig:compute_host_storage_usage_capacity} shows the percentage of free local storage per compute host. 
An uneven distribution of storage utilization is visible.
18\% of the host show more than 90\% of free storage, and 7\% are highly utilized requiring more than 30\% of storage.
While some hosts maintain either high or low utilization levels, other show more variability within the observed time frame.
Deriving clear trends needs additional data, which is currently not provided by our monitoring system.

\subsection{Workload Composition}
\label{sec:workload-composition}

Lastly, we analyze the composition of the workload within a region.

\paragraph{VM Resource Utilization}
We start by analyzing the CPU and memory utilization of the \glspl{vm}.
\autoref{fig:vm_resource_utilization_ratio} shows the cumulative distribution of CPU and memory utilization over 30 days over all \glspl{vm}.
We classify a \gls{vm} as underutilized when it consumes less than 70\%, optimally utilized between 70\% to 85\%, and overutilized when exceeding the 85\% threshold of the respective resources on average.
These thresholds are selected as \glspl{vm} using less than 70\% of allocated resources are considered inefficient, while those exceeding 90\% are seen as overutilized, based on our empirical observations and VMware best practice guidelines ~\cite[p.~26]{vmware_vsphere_80_perfomance_best_practices}.
Both potentially lead to resource allocation inefficiencies of the infrastructure.

\autoref{fig:vrops_virtualmachine_cpu_usage_ratio_all} depicts the average CPU utilization per \gls{vm} over the observation period.
The data indicates that most \glspl{vm} seem to be overprovisioned and consequently consume only a fraction of the allocated CPU resources.
Only a small set of the \gls{vm} workload seems to be optimally utilized.
An even smaller fraction can be classified as overutilized.

\autoref{fig:vrops_virtualmachine_memory_usage_ratio_all} presents the average memory utilization of the \gls{vm} workload.
In contrast to the CPU utilization, the memory indicates a different utilization pattern.
Approximately 38\% of \glspl{vm} seem to use less than 70\% of the requested memory and can be described as underutilized.
Around 10\% of \glspl{vm} can be classified as optimally utilized and a large proportion of \glspl{vm} consume more than 85\% of the allocated memory.

While multiple factors can influence scheduling, the overall resource allocations per \gls{vm} remain important.
\autoref{tab:vm_classification_vcpu} and \autoref{tab:vm_classification_ram} present a classification of \glspl{vm} based on the number of vCPU cores and RAM in GiB. 
The categories small, medium, large, and extra large define thresholds to distinguish lightweight from more resource-intensive \glspl{vm}.
Understanding and classifying the resource requirements of the workload can help to enable efficient resource and workload management by more effectively aligning the demand with the available infrastructure.

\paragraph{SAP Workloads}
SAP systems, such as the SAP S/4HANA, consist of multiple interconnected components including the SAP ABAP platform as the application server and SAP HANA in-memory databases.
Components of the application server infrastructure can be predominantly observed in the small, medium, and large categories.
The majority of SAP HANA in-memory database applications can be observed in the extra large category.

These systems include application-level resource management capabilities, which typically allocate and manage resources of the underlying \gls{vm}.
From a technical perspective, insight on the actual resource utilization, including caching behavior are not available to infrastructure providers and would require instrumentation on customer application level.
In addition to these technical constraints, details on the resource management at the SAP application level are considered sensitive data with business-critical implications.
However, workload characteristics on \gls{vm} level are representative, generalizable, and provided to the research community.

Additionally, workloads within the small, medium, and large categories consist of general purpose workloads consisting of a variety of applications, such as development environments, \gls{ci} and \gls{cd} systems, Kubernetes infrastructure.
The SAP Cloud Infrastructure offers \gls{iaas} capabilities and respects the confidentiality of the customers workloads. 
Thus, details beyond the current scope cannot be provided.

\begin{table}[h!]
    \centering
    \caption{Average of virtual machine classification based on the number of vCPUs}
    \label{tab:vm_classification_vcpu}
    \begin{tabular}{ p{2cm} r r p{2cm}}
        \hline
        \textbf{Category} & \textbf{vCPU (Cores)} & \textbf{Number of VMs} \\ \hline
        Small & $\leq 4$ & 28,446 \\ \hline
        Medium & $4 < \mathrm{vCPU} \leq 16$ & 14,340 \\ \hline
        Large & $16 < \mathrm{vCPU} \leq 64$ & 1831 \\ \hline
        Extra Large & $> 64$ & 738 \\ \hline
    \end{tabular}
\end{table}

\begin{table}[h!]
    \centering
    \caption{Average of virtual machine classification based on memory resources}
    \label{tab:vm_classification_ram}
    \begin{tabular}{ p{2cm} r r p{2cm}}
        \hline
        \textbf{Category} & \textbf{RAM (GiB)} & \textbf{Number of VMs} \\ \hline
        Small & $\leq 2$ & 991 \\ \hline
        Medium & $2 < \mathrm{RAM} \leq 64$ & 41,395 \\ \hline
        Large & $64 < \mathrm{RAM} \leq 128$ & 787 \\ \hline
        Extra Large & $> 128$ & 2184 \\ \hline
    \end{tabular}
\end{table}

\begin{figure*}%
    \centering
    \begin{subfigure}[b]{1\textwidth}
      \includegraphics[width=\textwidth]{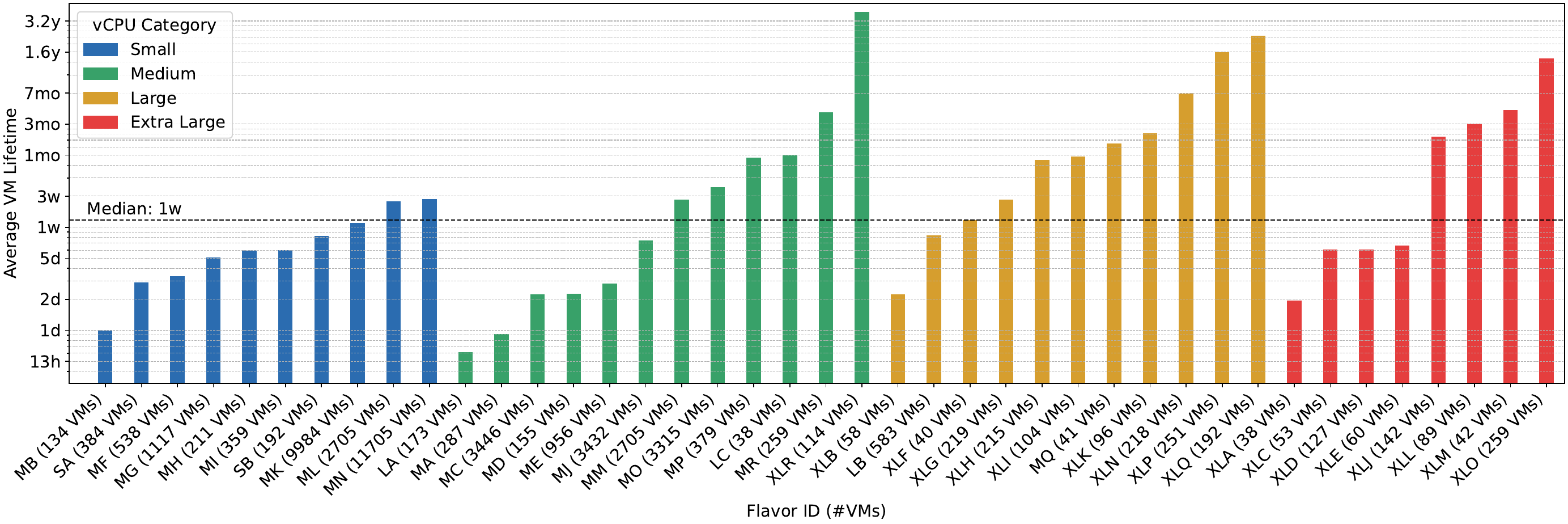}
      \caption{Categorized by vCPU class}
      \label{fig:vm_lifetime_vcpu}
    \end{subfigure}
    \begin{subfigure}[b]{1\textwidth}
      \centering
      \includegraphics[width=\textwidth]{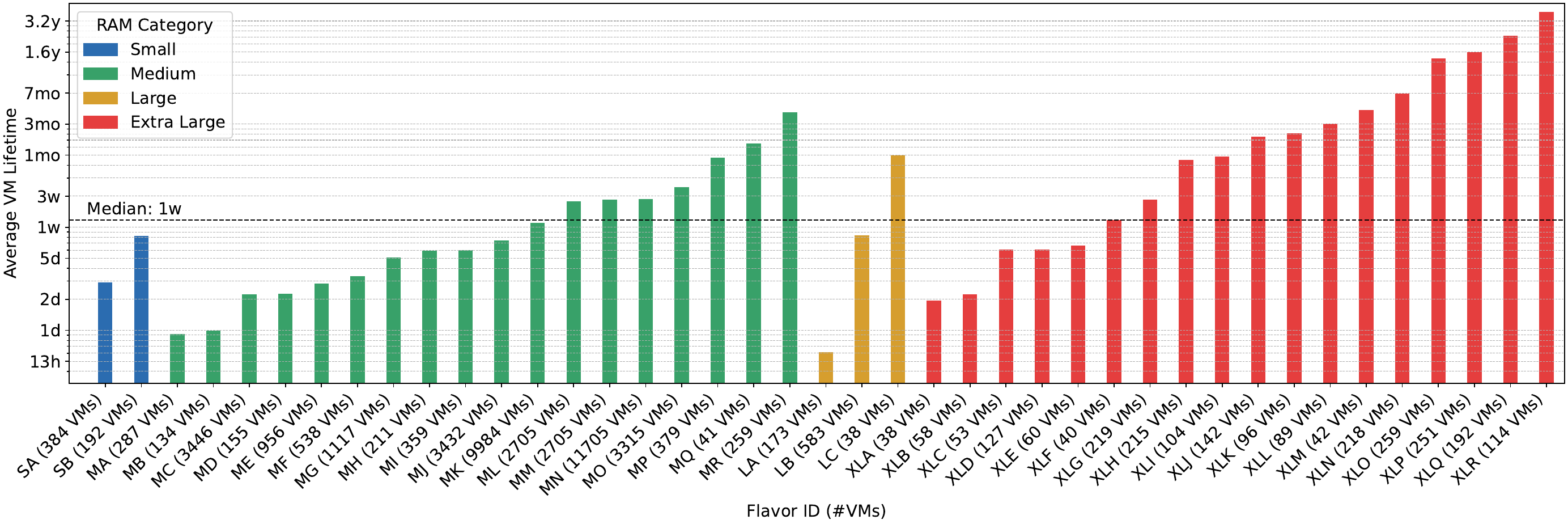}
      \caption{Categorized by RAM class}
      \label{fig:vm_lifetime_ram}
    \end{subfigure}
   \caption{VM lifetime per flavor grouped by different resource requirements in terms of processing and memory. The number of virtual machines instantiated per flavor are noted in braces.}
   \label{fig:vm_lifetime}
\end{figure*}

\paragraph{VM Lifetime}
The \gls{vm} lifetime is an important dimension, as it affects placement and scheduling decisions, consolidation opportunities, and the design of rebalancing systems \cite{10.1145/3669940.3707226}.
While short-lived \glspl{vm} can provide opportunities for dynamic load balancing, longer-lived \glspl{vm} can reduce the flexibility and contribute to resource fragmentation.

The \gls{vm} lifetime data was not available during the 30-day observation period.
We identified the gap and retrospectively collected the information from OpenStack.
As a result, this reflects only a post-observation state and does not represent the full temporal history.

\autoref{fig:vm_lifetime} illustrates the average lifetime of \glspl{vm} per flavor, categorized by vCPU and RAM class. 
Each flavor bar was annotated with the number of observed \glspl{vm} to provide an impression on the importance and distribution.
To avoid congestion, we limited the plots to flavors with at least 30 instances.
The observed lifetimes range from few minutes to multiple years and reflect the workload diversity of the SAP Cloud Infrastructure.
Some general-purpose and notably the memory-intensive flavors exhibit significant lifetimes and indicate stable and long-term deployments.

We observe significant variation in \gls{vm} lifetime within each category.
As a result, conclusions from \gls{vm} size to lifetime are limited.
Small \glspl{vm} do not consistently live shorter, nor large \glspl{vm} longer.

\section{Related Work}
\label{sec:related-work}

\renewcommand{\arraystretch}{0.4}
\begin{table*}
  \caption{Comparison of prior work and the SAP Cloud Infrastructure Dataset. The SAP dataset is the only publicly available dataset that provides VM workloads, including highly intensive memory allocations of up to 12TB per~VM, and a high sampling resolution covering nodes and VMs.}
  \label{tab:cloud_datasets_part1}
  \centering
  \begin{tabular}{ l ccccc ccr R{2.1cm} R{2.0cm} r r r }
    \toprule
    & \multicolumn{5}{c}{Resources} & \multicolumn{3}{c}{Workload} & \multicolumn{3}{c}{Vantage Points and Data Sampling} \\
    \cmidrule(rl){2-6}
    \cmidrule(){7-9}
    \cmidrule(lr){10-12}
    Dataset
          & \rotatebox[origin=c]{90}{CPU}
          & \rotatebox[origin=c]{90}{Memory}
          & \rotatebox[origin=c]{90}{Network}
          & \rotatebox[origin=c]{90}{Storage}
          & \rotatebox[origin=c]{90}{GPU}
          & \rotatebox[origin=c]{90}{Batch Jobs}
          & \rotatebox[origin=c]{90}{VMs}
          & \makecell[c]{\rotatebox[origin=c]{90}{Lifetime}}

          & \makecell[c]{\rotatebox[origin=c]{90}{Scale}}
  
          & \makecell[c]{\rotatebox[origin=c]{90}{Duration}}
          & \makecell[c]{\rotatebox[origin=c]{90}{Sampling}}
           & \rotatebox[origin=c]{90}{Public}

          \\
    \midrule

    Google \cite{google_cluster_usage_traces_v3}
          & \cmark & \cmark & \xmark & \xmark & \xmark
          & \cmark & \xmark
          & sec–days
          & 672,074 jobs
          & 29 days
          & 5 min
          & \cmark

          \\

    Alibaba \cite{alibaba_cluster_trace_v2018}
          & \cmark & \cmark & \xmark & \cmark & \cmark
          & \cmark & \xmark
          & min–days
          & $\sim$4k nodes
          & 8 days
          & n/a 
          & \cmark

          \\

    Philly \cite{jeon2019analysislargescalemultitenantgpu}
          & \cmark & \cmark & \cmark & \xmark & \cmark
          & \cmark & \xmark
          & min–weeks
          & 117,325 jobs
          & 75 days
          & 1 min 
          & \cmark

          \\
	
    Atlas \cite{amvrosiadis2018atlas}
          & \cmark & \cmark & \cmark & \xmark & \cmark
	    & \cmark & \xmark
	    & n/a

	    & 96,260 jobs
	    & 90--1,800 days
	    & 1 min 

          & \cmark

          \\

    MIT \cite{Samsi_2021}
          & \cmark & \cmark & \cmark & \cmark & \cmark
          & \cmark & \xmark
          & min–days
          & 441--9k nodes
          & 90--180+ days
          & n/a %
          & \cmark

          \\
	
    \midrule

    Azure \cite{10.1145/3669940.3707226}
          & \cmark & \cmark & \cmark & \cmark & \xmark
          & \xmark & \cmark
          & min–weeks
          & $>$1M VMs
          & 14 days
          & 5 min 
          & \xmark

          \\

    \dhline

    SAP (this work)
          & \cmark & \cmark & \cmark & \cmark & \xmark
          & \xmark & \cmark
          & min–years
          & 1.8k nodes, 48k VMs
          & 30 days
          & 30s--300s 
          & \cmark

          \\
          
    \bottomrule
  \end{tabular}
\end{table*}

Prior work introduced datasets from commercial cloud datasets.
Samsi \etal~\cite{Samsi_2021} provide an overview of several publicly available datasets from Azure (Philly trace), Google, Blue waters, Atlas, and others.
The datasets span various time frames ranging from short-term collections of 7 hours to extensive data collections over 80 months.
They focus primarily on CPU, GPU, memory, network, storage I/O, latency, temperature, and power consumption metrics.
These datasets enabled various researchers to analyze and address resource utilization, scheduling, and workload management.

In addition, Samsi \etal\cite{Samsi_2021} present the MIT Supercloud dataset covering more than 6 months of scheduling-relevant data on node, CPU, memory, hardware utilization, and filesystem I/O performance.
The dataset stands out with its level of detail, containing data points every 100 ms for GPU metrics, 10s for CPU metrics, and latency samples every 5 min.
The authors present the data to advance \gls{ai} and \gls{ml} research on workload optimization, predictive maintenance, and resource efficiency.

Publicly available traces from the Alibaba cloud \cite{alibaba_clusterdata} include data from 2017, 2018, 2020, and 2021.
The datasets consist of various metrics and focus on machine data, CPU and GPU cluster utilization, memory bandwidth, and resource contention indicators.
While some datasets are described to have workloads, others have specific applications and usage scenarios.
The largest dataset covers up to 1800 machines over the course of 2 months with \gls{ai} and \gls{ml} workload.
Several technical reports and research publications were based on the provided dataset aiming to enable research based on realistic data and improve workload characterization, optimize resource efficiencies, and scheduling \cite{288717, 10.1145/3545008.3545026, 10.1145/3326285.3329074}.

Reidys \etal \cite{10.1145/3669940.3707226} propose Coach, a system that exploits temporal resource patterns in \gls{vm} resource utilization to improve resource efficiency.
Motivated by this work, we collected \gls{vm} lifetime data, as such metrics were not consistently available in our infrastructure.

While the mentioned studies provide important insights, our work differs from them in several aspects. 

Traces, such as Google~\cite{google_cluster_usage_traces_v3} and Alibaba~\cite{alibaba_clusterdata}, predominantly capture short-lived, batch-oriented, and \gls{ai}\/\gls{ml}-intensive workloads in containerized settings.
In contrast, our dataset presents long-lived, memory-intensive, and \gls{vm}-based SAP enterprise workloads.
The difference in workload types can lead to different resource characteristics.
Existing traces highlight frequent task scheduling events, while our dataset reveals persistent resource contention and extended \gls{vm} lifetimes.

For our resource-intensive, long-lived, and stateful workloads (\eg in-memory databases), availability and memory residency take precedence over elasticity.
Thus, they may require different placement and scheduling mechanisms than transient and stateless tasks.
The development of such mechanisms benefits from traces exhibiting these characteristics, which our dataset provides.

Additionally, our dataset differs from the existing studies by providing comprehensive and fine-grained telemetry data on both hypervisor and \gls{vm} level.
This includes CPU, memory, storage, network, and lifetime metrics, which enable a more comprehensive analysis across different resources.

We summarize our discussion of related work in \autoref{tab:cloud_datasets_part1}.

This study complements existing studies by presenting empirical insights from production workloads of the SAP Cloud Infrastructure and extends workload characterization to enterprise environments that have so far received limited attention.

\section{Discussion and Guidance}
\label{sec:discussion}

Resource contention is a critical problem in cloud computing environments because it can negatively impact the performance and efficiency of the workload.
It is imperative to prevent performance bottlenecks and ensure stability of the platform.
This begins with an understanding and systematic characterization of the environment and its workload.
We now contextualize our results to inform strategies to mitigate potential suboptimal scheduling situations.

\paragraph{Placement and dynamic rescheduling mechanisms should be combined}
One of the key findings is that the resource utilization over most compute nodes is relatively static within the considered time frame.
Our data shows overutilization and resource contention situations on some nodes while others remain underutilized.
Dynamic rescheduling capabilities can continuously redistribute workloads and thus mitigate fragmentation caused by the initial placement of workloads in clustered infrastructures.
Our observations indicate that combining placement decisions with dynamic rescheduling mechanisms may help to achieve more balanced utilization. 
Such a unified, ideally even proactive, approach may also reduce the number of required workload migrations, each of which can incur costs in the form of temporary performance impacts.

\paragraph{Overprovisioning is still common}
Multiple studies \cite{huan_server_utilization, 10.1145/2391229.2391236, 10.1145/2644865.2541941} show that users tend to overprovision \glspl{vm} by allocating more resources than they actually utilize over the course of their lifetime.
Our data suggests that this is still the case in particular for CPU resources.
Whereas CPU resources are significantly overprovisioned, memory requests seem to be better aligned with the actual usage.
This raises the question to which extent the CPU contention on the compute hosts impairs workload performance.
Application-level insights would be required to address this question.
Gaining those insights is challenging, though, because of limited monitoring~capabilities.

Potential guidance could be twofold.
First, the overcommit factor should be reconsidered.
Infrastructure providers often split physical cores~(pCPU) into multiple virtual cores~(vCPU). 
To simplify provisioning, they calculate the overcommit factor as a ratio of vCPUs to pCPUs.
A more dynamic and workload-based approach to determine the overcommit factor and related configuration might help to mitigate these problems and improve the resource utilization of the overall landscape.
Several key metrics (see \autoref{sec:optimization-criteria}) can help to assess the overcommit factor.

Second, recommendations about qualified right-sizing could help users to adjust the requested resources and the associated costs based on the actual usage.

\paragraph{Common software platforms allow for extension, which should be leveraged}
The OpenStack Nova scheduler has by default limited placement capabilities and solely relies on current data.
Deployments should make use of the option to extend the scheduler.
Enhancements to the initial placement capabilities could be a potential area for exploration.
These might involve incorporating both current and historic utilization data, for example the contention metrics.

From an architectural perspective, the placement and scheduling of \glspl{vm} occur independently on two layers. 
First, the Nova scheduler assigns a \gls{vm} to a vSphere cluster and subsequently the VMware \acrlong{drs} allocates the \gls{vm} to a specific host.
This independent scheduling can cause fragmentation and suboptimal scheduling especially in environments with isolated vSphere clusters.
A holistic scheduler that assigns \glspl{vm} directly to individual hosts might be capable of improving resource utilization and reduce fragmentation.

\paragraph{Imbalances caused by infrastructure fragmentation should be addressed}
Fragmentation across logically grouped resources, such as \glspl{bb}, results in measurable imbalances that impair scheduling efficiency and reduce infrastructure utilization.
Continuous migration mechanisms across \glspl{bb} are required to maintain balanced resource distribution.

\paragraph{Workload lifetime contributes to resource fragmentation}
Workload lifetime can influence resource fragmentation.
Long-lived \glspl{vm} occupy resources for extended periods, which may increase fragmentation.
Placement strategies that incorporate workload lifetime can reduce migrations and mitigate resource fragmentation.
This can help to improve scheduling efficiency and infrastructure utilization.

\paragraph{Holistic scheduling across layers is required}
Independent scheduling mechanisms across multiple infrastructure layers can lead to local optimizations but global inefficiencies.
A holistic approach could reduce fragmentation and improve system-wide resource management.

\paragraph{CPU contention should be mitigated}
CPU contention can occur when co-located workloads exceed defined CPU utilization thresholds.
In such situations, workload performance may be impaired.
These observations motivate further analysis of possible root causes, such as noisy-neighbor situations or time-synchronous events.
They further motivate the use of percentage-based scheduling triggers, since absolute thresholds should be avoided due to differences in infrastructure and hypervisor implementations, and contention-aware algorithms to mitigate potential effects.

\paragraph{Memory-based bin-packing strategies are required}
A subset of the workload presented in this study is memory-bound.
In these cases, memory-based bin-packing strategies should be applied to optimize resource utilization and avoid inefficiencies.
Moreover, this underlines the importance of workload characterization as a prerequisite for selecting appropriate bin-packing strategies.

\section{Conclusion and Outlook}
\label{sec:conclusions}

Placement and scheduling of \glspl{vm} are critical challenges in achieving efficient resource utilization in large-scale cloud computing environments.
In this work, we highlighted the importance of a data-driven approach to address these challenges.
We analyzed a real-world dataset of the SAP Cloud Infrastructure, consisting of 1,800 hypervisors and 48,000 \glspl{vm}.
Our findings reveal inefficiencies in scheduling with vanilla OpenStack, including shortcomings in workload-specific scheduling, and resource contention situations, which lead to suboptimal resource utilization.
Moreover, we provide insights into characteristics of a diverse workload.

By offering a publicly available dataset of a real-world cloud computing environment, we aim to advance research in the area of placement and scheduling in large-scale cloud environments.
The insights can help evaluate existing solutions and inform the development of novel scheduling solutions for large-scale cloud environments.

While our dataset covers various aspects of scheduling, we plan to add additional metrics such as performance of \glspl{vm} and hypervisors, and the number of \gls{vm} migrations.
Furthermore, \gls{qos} requirements provide guarantees for certain performance standards such as latency, network bandwidth, disk I/O, \gls{numa} alignment, and CPU-pinning.
The latter ensures reduced latency to performance-sensitive \glspl{vm} by reserving dedicated CPU cores on hosts.
In our future work, we plan to evaluate OpenStack \gls{qos} classes for more fine-grained management of different types of \glspl{vm}.

Future work should extend the analysis beyond the 30-day observation period and the single regional deployment studied here.  
Longer observation windows could capture seasonal workload variations, while studies across multiple regions could reveal structural differences in scheduling efficiency.

We hope that our data-driven experiences will contribute to a better understanding and advance research on placement and scheduling in large-scale cloud environments.

\begin{acks}
    We would like to thank the anonymous reviewers and our shepherd for their feedback on this paper, as well as the following individuals for their support (in alphabetical order): Ralf Ackermann, Philipp~Matthes, and Michael Schmidt.
    We gratefully acknowledge SAP for supporting open science and reproducible research.

    This study was carried out as part of the project ``Apeiro Reference Architecture'', in short ApeiroRA, under the grant \grantnum{EU}{13IPC007}, partly financed by the \grantsponsor{EU}{European Union -- NextGenerationEU}{https://next-generation-eu.europa.eu/} and partly funded by the \grantsponsor{BMWE}{German Federal Ministry for Economic Affairs and Energy}{https://www.bundeswirtschaftsministerium.de/Navigation/EN/}.

\end{acks}


\label{lastpagebody}

\bibliographystyle{ACM-Reference-Format}
\bibliography{res/sources}

\appendix
\makeatletter
\if@twocolumn
  \if@firstcolumn
    \newpage
  \fi
\fi
\makeatother
\begingroup\raggedbottom

\section{Ethics}
\label{sec:ethics}

This work does not raise any ethical issues.
The data presented in this publication is fully anonymized and does not include personally identifiable information.
Metadata, such as hostnames, project IDs, and IP addresses were consistently hashed or removed.

\section{Dataset}
\label{sec:dataset}

The full dataset used in this study is made publicly available.\newline
\textbf{Dataset}: Anonymized telemetry data in CSV format.\newline
\textbf{Run-time environment}: Python\newline
\textbf{Publicly available?} Yes\newline
\textbf{Archived?} Yes, at \href{https://doi.org/10.5281/zenodo.17141306}{https://doi.org/10.5281/zenodo.17141306} ~\cite{zenodo17141306}

\section{Metrics}
\label{sec:metrics}

\autoref{tab:metrics} provides an overview of resource utilization metrics collected from VMware vROps and OpenStack Compute.
The metrics include data on CPU, memory, network, and storage across compute hosts and \glspl{vm} within a region.

\section{Details About SAP Data Centers}
\label{sec:dcs-details}

\autoref{tab:cloud_numbers} presents the distribution of hypervisors and virtual machines across SAP's globally distributed data centers.

\begin{table}[H]
  \centering
  \caption{Metric details for vROps and OpenStack Compute}
  \label{tab:metrics}
  \setlength{\tabcolsep}{5pt}\renewcommand{\arraystretch}{1.15}
  \begin{adjustbox}%
    {angle=90,center, totalheight=\dimexpr\textheight-\abovecaptionskip-\belowcaptionskip-2ex\relax, max width=\columnwidth, keepaspectratio}
    \begin{tabular}{ | l | c | c | c | }
      \hline
      Metric Name & Subsystem & Resource & Description \\
      \hline
      vrops\_hostsystem\_cpu\_core\_utilization\_percentage & Compute host & CPU & Utilization of CPU per compute host \\
      vrops\_hostsystem\_memory\_usage\_percentage & Compute host & Memory & Utilization of compute host memory \\
      vrops\_hostsystem\_network\_bytes\_rx\_kbps & Compute host & Network & Received network traffic \\
      vrops\_hostsystem\_network\_bytes\_tx\_kbps & Compute host & Network & Transmitted network traffic \\
      vrops\_hostsystem\_diskspace\_usage\_gigabytes & Compute host & Storage & Utilization of local storage \\
      vrops\_hostsystem\_cpu\_contention\_percentage & Compute host & CPU & Observed CPU contention per compute host \\
      vrops\_hostsystem\_cpu\_ready\_milliseconds & Compute host & CPU & Duration a VM is ready but waits for scheduling \\
      vrops\_virtualmachine\_cpu\_usage\_ratio & VM & CPU & Percentage of requested and used CPU \\
      vrops\_virtualmachine\_memory\_consumed\_ratio & VM & Memory & Percentage of requested and used memory \\
      \hline
      openstack\_compute\_instances\_total & Region & – & Total number of VMs within the regional deployment \\
      openstack\_compute\_nodes\_vcpus\_gauge & Compute host & CPU & Number of vCPUs per compute host \\
      openstack\_compute\_nodes\_vcpus\_used\_gauge & Compute host & CPU & Number of vCPUs per compute host \\
      openstack\_compute\_nodes\_memory\_mb\_gauge & Compute host & Memory & Amount of memory in MB per compute host \\
      openstack\_compute\_nodes\_memory\_mb\_used\_gauge & Compute host & Memory & Amount of utilized memory in MB per compute host \\
      \hline
    \end{tabular}
  \end{adjustbox}
\end{table}
\FloatBarrier

\begin{table}[H]
  \centering
  \caption{Data center overview}
  \label{tab:cloud_numbers}
  \small
  \begin{filecontents*}{res/dc_numbers.csv}
Region ID,Datacenter name,Number of Hypervisors,Number of VMs
1,A,167,4985
1,B,65,375
2,A,244,7913
2,B,112,1284
3,A,202,4475
3,B,89,1353
4,A,191,3977
5,A,42,395
6,A,150,5016
7,A,63,1096
8,A,227,5595
8,B,270,4206
8,D,966,34392
9,A,751,19464
9,B,1072,27652
10,A,65,1186
10,B,152,5713
11,A,60,2877
12,A,62,1996
12,B,43,362
13,A,274,7432
13,B,99,1149
13,D,239,3881
14,A,330,3809
14,B,307,5125
15,A,209,5442
16,A,40,504
16,B,28,156
16,D,22,78
\end{filecontents*}
\csvreader[tabular=|c|c|c|c|,
    table head=\hline
      \rule{0pt}{2em} %
      \textbf{\shortstack{Region \\ ID}} & 
      \textbf{\shortstack{Datacenter \\ Name}} & 
      \textbf{\shortstack{Number of \\ Hypervisors}} & 
      \textbf{\shortstack{Number of \\ Virtual Machines}} \\ \hline,
    late after line=\\\hline,
    before reading={\renewcommand{\arraystretch}{1.5}}]
  {res/dc_numbers.csv}{}%
  {\csvlinetotablerow}
\end{table}
\FloatBarrier

\label{lastpage}

\end{document}